\documentclass{article}
\usepackage{amssymb}
\usepackage{amsmath}
\usepackage[dvips]{graphicx}
\usepackage{float}
\usepackage{afterpage}
\usepackage{eufrak}

\begin{document}
  \title{Consequences of Dirac's Theory of the Positron}
  \author{W. Heisenberg and H. Euler in Leipzig$^1$}
  \date{22. December 1935}
  \maketitle

\begin{abstract}
According to Dirac's theory of the positron,
an electromagnetic
field tends to create pairs of
particles which leads to
a change of Maxwell's
equations in the vacuum. These changes are
calculated in the special case that no real electrons or
positrons are present and the field
varies little over a Compton wavelength.
The resulting effective Lagrangian of the field reads:
\begin{equation*}
\begin{split}
\scriptstyle
\mathfrak{L}=
& \scriptstyle \frac{1}{2}(\mathfrak{E}^2 -\mathfrak{B}^2) + \frac{e^2}{\hbar c}
\int_0^{\infty} e^{-\eta}\frac{d\eta}{\eta ^3}\biggl\{\scriptstyle i\eta ^2 (\mathfrak{EB}) \cdot
\frac{\ \cos{(\frac{\eta}{|\mathfrak{E}_k|} \sqrt{\mathfrak{E}^2-\mathfrak{B}^2+2i(\mathfrak{EB})})} + \text{conj.}}
{\cos{(\frac{\eta}{|\mathfrak{E}_k|} \sqrt{\mathfrak{E}^2-\mathfrak{B}^2+2i(\mathfrak{EB})})} - \text{conj.}}\\
& \qquad \qquad \qquad \qquad \qquad \qquad \qquad \qquad \qquad
\scriptstyle +|\mathfrak{E}_k | ^2+\frac{\eta ^2}{3}(\mathfrak{B}^2-\mathfrak{E}^2)
\biggr\}\\
& \qquad \qquad \scriptstyle \mathfrak{E},\mathfrak{B}\quad \text{ field strengths}\\
& \qquad \qquad \scriptstyle |\mathfrak{E}_k |= \frac{m^2 c^3}{e \hbar}=\frac{1}{137}\frac{e}{(e^2/m c^2)^2} =\quad \text{ critical field strengths}\\
\end{split}
\end{equation*}
The expansion terms in small fields (compared to $\mathfrak E$ ) describe light-light scattering.
The simplest term is already known from perturbation theory.
For large fields, the equations derived here
differ strongly from
Maxwell's equations. Our  equations
will be compared to those proposed by Born.
\\ [3mm]
German title:
``Folgerungen aus der Diracschen Theorie des Positrons"
Zeitschr. Phys. {\em 98\/}, 714 (1936). \\[2mm]
Translation:
W. Korolevski and H. Kleinert,
Institut f\"ur Theoretische Physik,
Freie Universit\"at Berlin, Arnimallee 14, D-14195 Berlin, Germany. \\
emails: {\tt walja.k@web.de} and {\tt kleinert.physik.fu-berlin.de}
\end{abstract}
The fact that electromagnetic radiation can be transformed into matter and vice versa leads to
fundamentally new features in quantum electrodynamics.
One of the most important consequences is that, even in the vacuum, the
Maxwell equation have to be exchanged by more complicated formulas. In general,
it will be not possible to separate processes in the vacuum from those involving matter
since electromagnetic fields can create matter if they are strong enough. Even if they
are not strong enough to create matter they will, due to the virtual possibility
of creating matter, polarize the vacuum and therefore change the Maxwell equations.
This polarization of the vacuum to be studied below
will give rise to a
distinction between the vectors $\mathfrak B$, $\mathfrak E$
on the one hand and $\mathfrak D$, $\mathfrak H$ on the other, where%

 \includegraphics[scale=0.6]{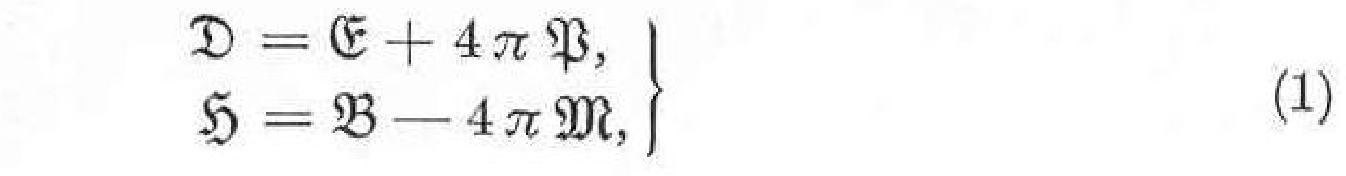}
The polarizations $\mathfrak P$ and $\mathfrak M$ can be arbitrary functions of
the field strengths at the same place, their derivatives, and the field strengths in the
surroundings of the observed position. If the field strengths are small
(which means, as we shall see small compared to $e^2/\hbar c$-times of the field strength
at the boundary of the electron), then $\mathfrak P$ and $\mathfrak M$ can be approximately
considered as linear functions of $\mathfrak E$ and
$\mathfrak B$. In this approximation, Uehling\footnote{E. A. Uehling, Phys. Rev. {\bfseries 48}, 55, 1935. \label{Ueh}}
and Serber\footnote{R. Serber, Phys. Rev. {\bfseries 48 }, 49, 1935. \label{Serber}} have calculated
the modifications of Maxwell's theory. Another interesting case is obtained by not assuming
small field strengths but instead slowly varying fields (i.e., the fields are nearly constant
over the length $\hbar/mc$).
Then one obtains $\mathfrak P$ and $\mathfrak M$ as functions of $\mathfrak E$
and $\mathfrak B$ at the same position. The derivatives of $\mathfrak E$ and $\mathfrak B$ do not appear
in the approximation. The expansion of $\mathfrak P$ and $\mathfrak M$ in powers of $\mathfrak E$ and $\mathfrak B$
will contain only odd powers, as will be seen in the calculation. The
expansion terms of
third order are
phenomenologically related to light-light scattering
and are already known\footnote{H. Euler, B. Kockel, Naturwissensch. {\bfseries 23}, 246, 1935.\label{EulerKockel}}.
The goal of this work is to find the functions
$\mathfrak P(\mathfrak E,\mathfrak B)$ and $\mathfrak M(\mathfrak E,\mathfrak B)$
for slowly varying
field strength. It is sufficient to calculate to energy density $U(\mathfrak E,\mathfrak B)$.
From the energy density one can derive the
fields using the Hamiltonian method; one introduces the
Lagrangian $\mathfrak L(\mathfrak E,\mathfrak B)$ and obtains
\includegraphics[scale=0.6]{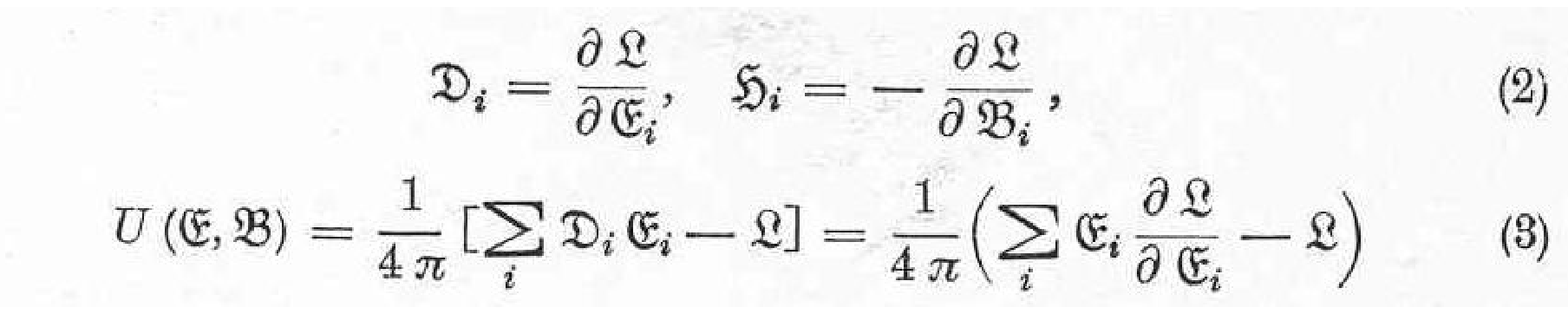}
The Lagrangian is determined by (3) and $\mathfrak D$ and $\mathfrak H$ are determined by (2).
Due to relativistic invariant the Lagrangian can only depend
on the two invariants ${\mathfrak E}^2-{\mathfrak B}^2$ and $({\mathfrak EB})^2$ (compare \ref{EulerKockel}).
The calculation of $U(\mathfrak E,\mathfrak B)$ can be reduced to the question of how much energy density is associated with the matter fields in a background of constant fields $\mathfrak E$ and
$\mathfrak B$. Before solving this problem, the mathematical scheme of the positron
theory\footnote{W. Heisenberg, ZS. f. Phys. {\bfseries 90}, 209, 1934\label{Heisenberg}} will be presented. This will also correct some errors in earlier formulas.

\section{The Mathematical Schema of the Theory of the Positron}

Starting point of the theory is the Dirac density matrix, which is given
in naive
wave theory by
\begin{equation}
(x' t' k' |R |x'' t'' k'') =   \underset{\scriptscriptstyle \text{ occupied states}}{\sum (n)}
\psi _n ^{*} (x''t''k'') \: \psi _n (x't'k')\tag{4}
\end{equation}
and in this quantum theory of wave fields by\footnote{In the earlier paper [\ref{Heisenberg}] the values on the right hand side are interchanged}
\begin{equation}
(x' t' k' |R |x'' t'' k'') =  \psi  ^{*} (x''t''k'') \: \psi (x't'k')\tag{5}
\end{equation}
Apart from this matrix an important role is played by the matrix $R_S$,
formally defined by
\begin{equation}
(x' t' k' |R_S |x'' t'' k'') = \frac{1}{2} \biggl (  \underset{\scriptscriptstyle \text{occupied states}}{\sum (n)} -
\underset{\scriptscriptstyle \text{unoccupied states}}{\sum (n)}\quad  \biggr )
\psi _n ^{*} (x''t''k'') \psi _n (x't'k')\tag{6}
\end{equation}
and in the quantum theory by
\begin{equation}
(x' t' k' |R_S |x'' t'' k'') = \frac{1}{2} \biggl[ \psi  ^{*} (x''t''k'') \: \psi (x't'k') -
\psi (x't'k') \: \psi ^* (x''t''k'') \biggr] \tag{7}.
\end{equation}
The matrix $R_S$ as a function of the differences $x_\lambda' - x_\lambda'' = x_\lambda$
and $t'-t''=t$ becomes singular on the light cone. If we set

\includegraphics[scale=0.6]{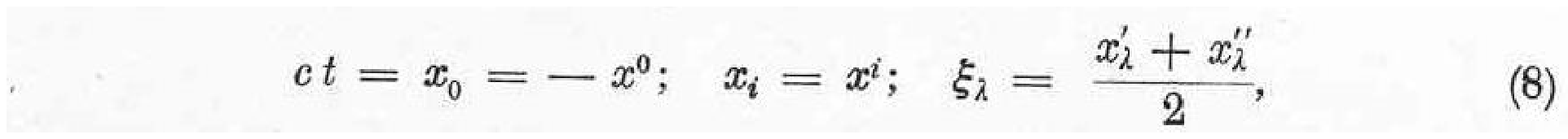}
furthermore the potentials $A_0=-A^0$; $A_i=A^i$, and  for the Dirac matrices
$\alpha^0=-\alpha_0=1$; $\alpha^i=\alpha_i$, we derive

\includegraphics[scale=0.6]{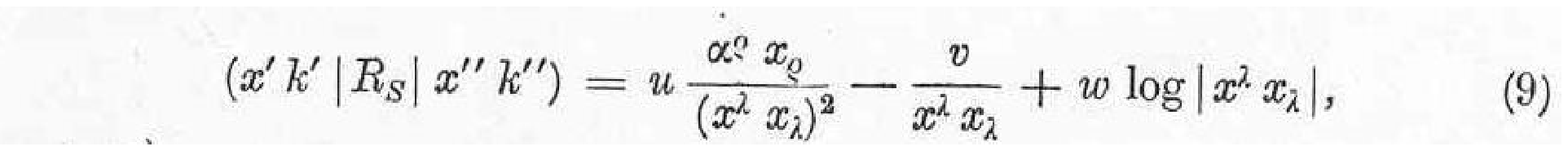}
where\footnote{In the earlier paper [\ref{Heisenberg}] the exponent has wrongly a negative sign.}

\includegraphics[scale=0.6]{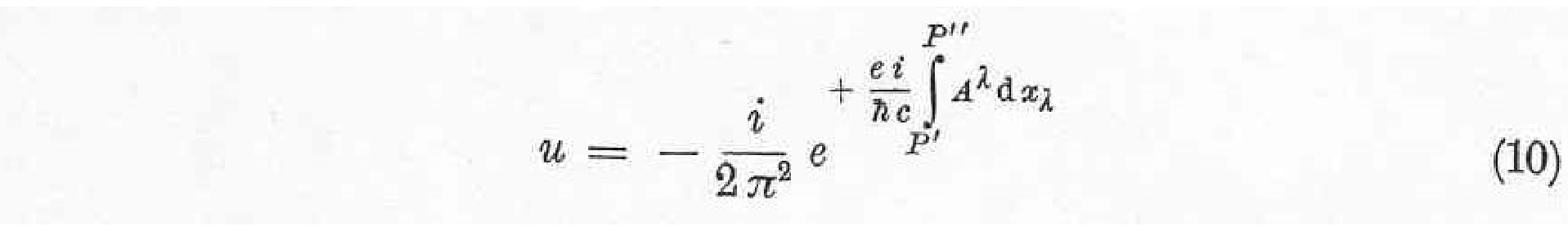}
(The sum of repeated Latin indices runs from 1 to 3, and of Greek indices from 0 to 3).
The integral is to be calculated along the straight line from $P'$ to $P''$.

The density matrix $r$ responsible for the behavior of matter is obtained from $R_S$ by the equation

\includegraphics[scale=0.6]{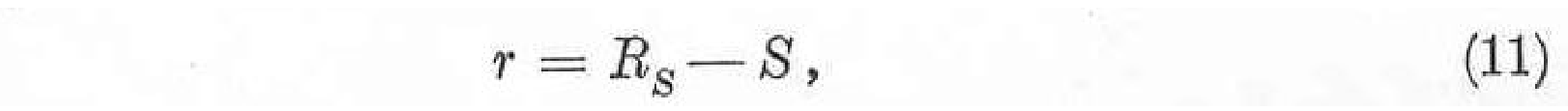}
where $S$ is given by

\includegraphics[scale=0.6]{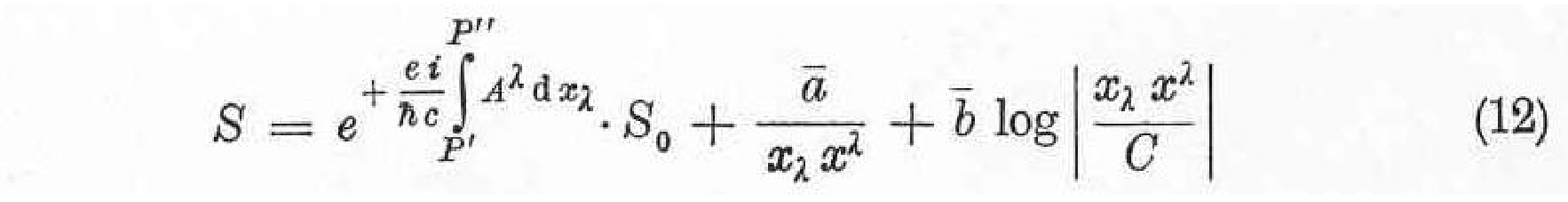}
Here $S_0$ is the matrix $R_S$ in field-free and matter-free space, where
$\bar{a}$, $\bar{b}$, and $C$ are defined\footnote{In the earlier paper [\ref{Heisenberg}] there is a mistake that leads to another value of C. There, the character $\gamma$ stands for the logarithm of the Euler constant, in contrast to the common usage.} by

\includegraphics[scale=0.6]{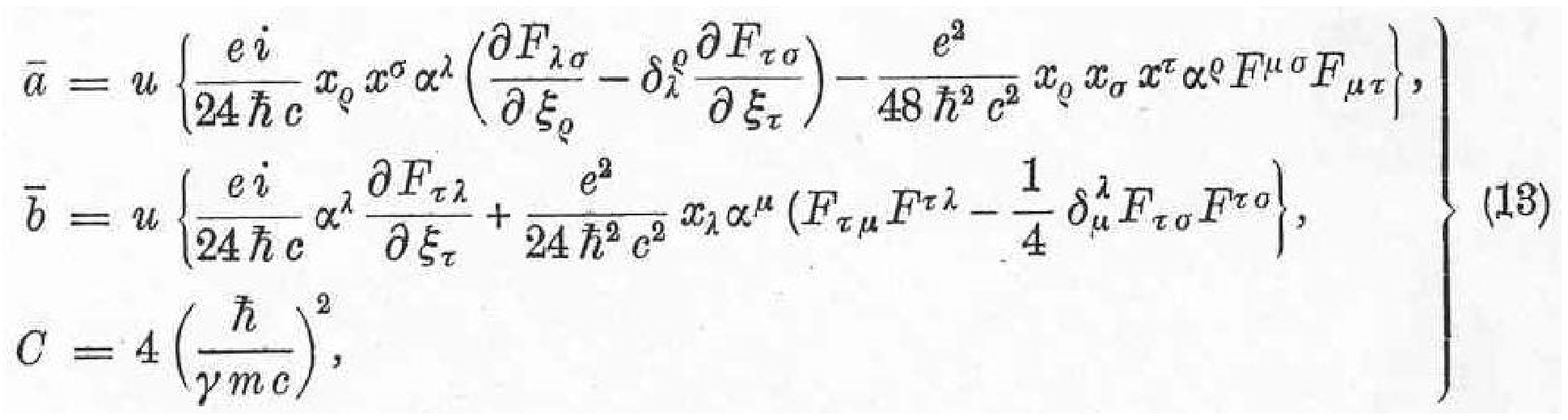}
where $\gamma$ is the Euler constant $\gamma=1.781...$

The four-vector for the current density and the energy momentum tensor are obtained from $r$ by

\includegraphics[scale=0.6]{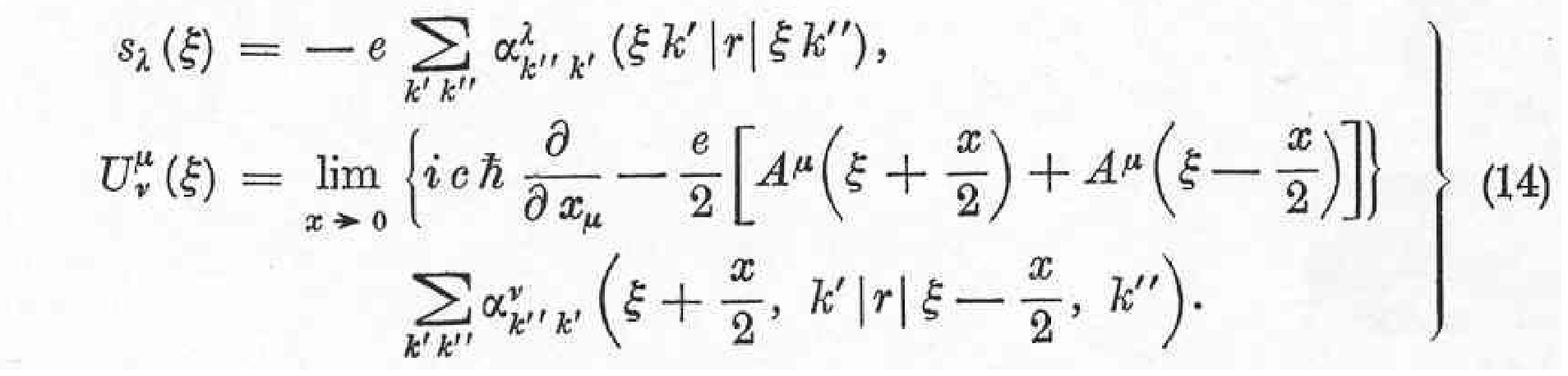}
In the quantum theory of wave fields it is useful to expand the wave function into an
orthogonal system:

\includegraphics[scale=0.6]{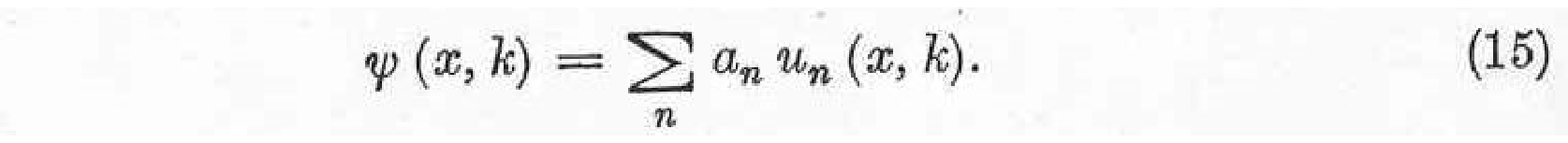}
The operators $a_n$ can be represented in the form

\includegraphics[scale=0.6]{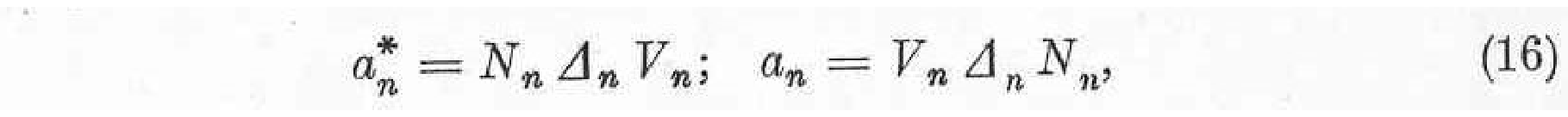}
where $\Delta_n$ changes the number $N_n$ to $1-N_n$ and
$V_n=\prod_{t\leq n}(1-2N_t)$.

We further set

\includegraphics[scale=0.6]{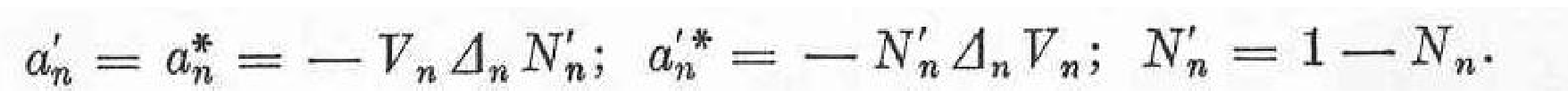}
The Hamiltonian of the whole system reads in these variables:

\includegraphics[scale=0.6]{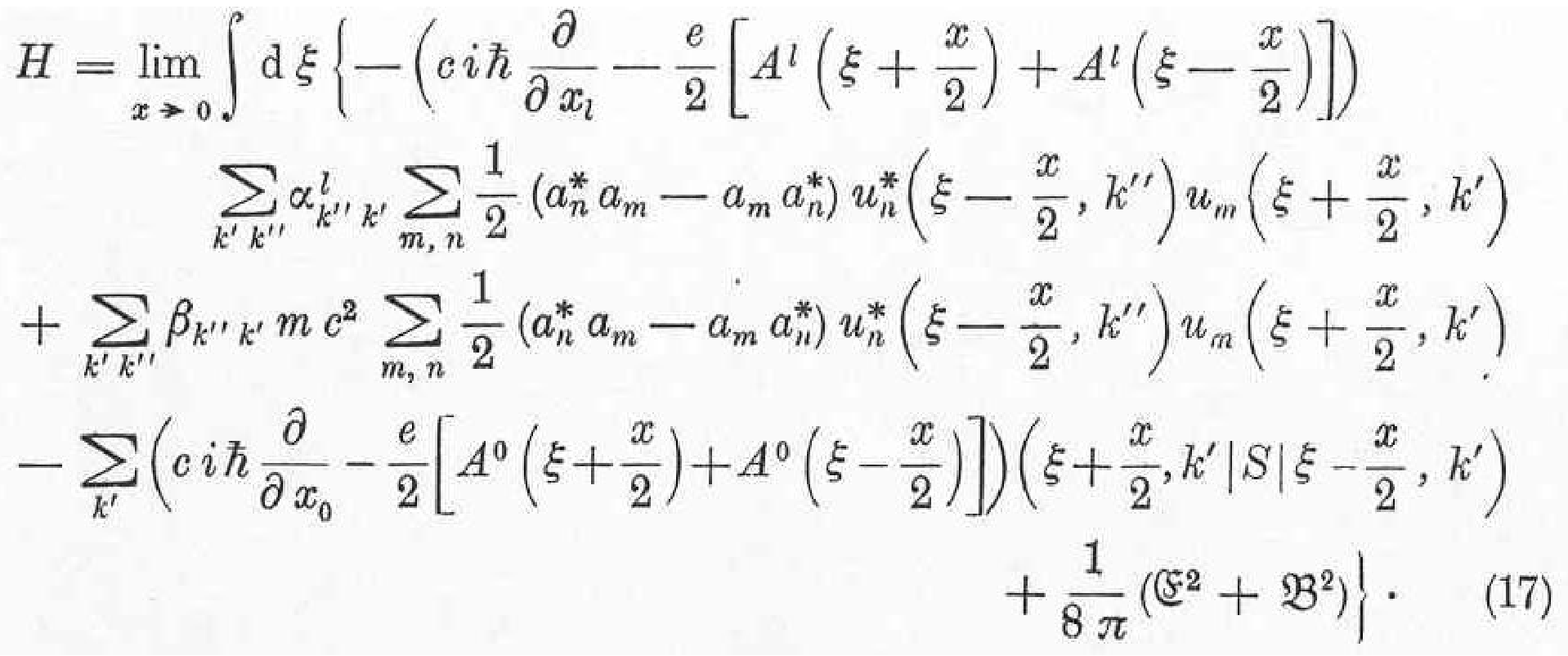}
According to the powers of the elementary change $e$, we distinguish the terms:

\includegraphics[scale=0.6]{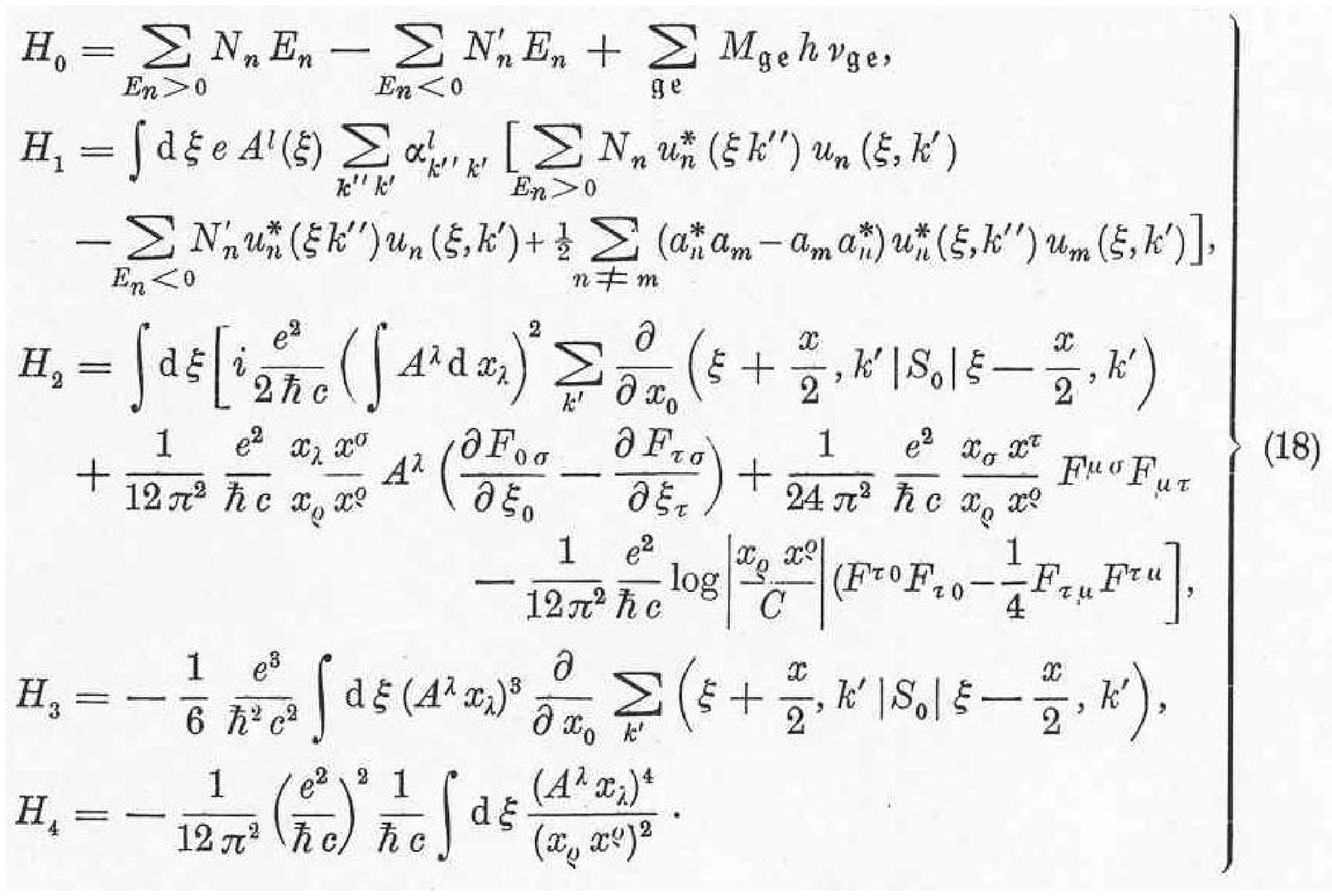}

\section{Calculation of the Energy Density in Intuitive Wave Theory}

Since the Lagrangian associated with the corrected Maxwell equations
is a function of the two invariants ${\mathfrak E}^2-{\mathfrak B}^2$ and $({\mathfrak EB})^2$,
it is sufficient to calculate the energy density of the matter field as a
function of two independent field quantities. For example, it will be sufficient
to calculate the energy density in a constant electric and a
constant parallel magnetic field. In these constant fields we have to
analyze the state of the matter field in the absence of matter,
i.e., in the state of lowest energy. In an intuitive wave theory based on equations (4) and (6),
the state of lowest energy is given when all electron states of negative energy
are occupied and all states of positive energy are empty. In the presence of only
a magnetic field, the stationary states of an electron can be divided into
those of negative and positive energy. Hence the state of the lowest energy
of the matter field can be derived in the same way as for a field-free space.

The situation is different in an electric field. In this case the potential
energy grows linearly in one space direction, so all energy values from $-\infty$ to $+\infty$
are possible. The eigenfunctions associated with different eigenvalues are transformed
into each other by a spatial translation. Hence a classification of energy values into
positive and negative is not unique.

This difficulty is physcially related to the fact that in an electric field,
pairs of positrons and electrons are created. The exact analysis of this problem
was performed by Sauter\footnote{F. Sauter, ZS. f. Phys. {\bfseries 69}, 742, 1931 \label{Sauter}}.

\begin{figure}[h]
        \centering
        \includegraphics[scale=0.3]{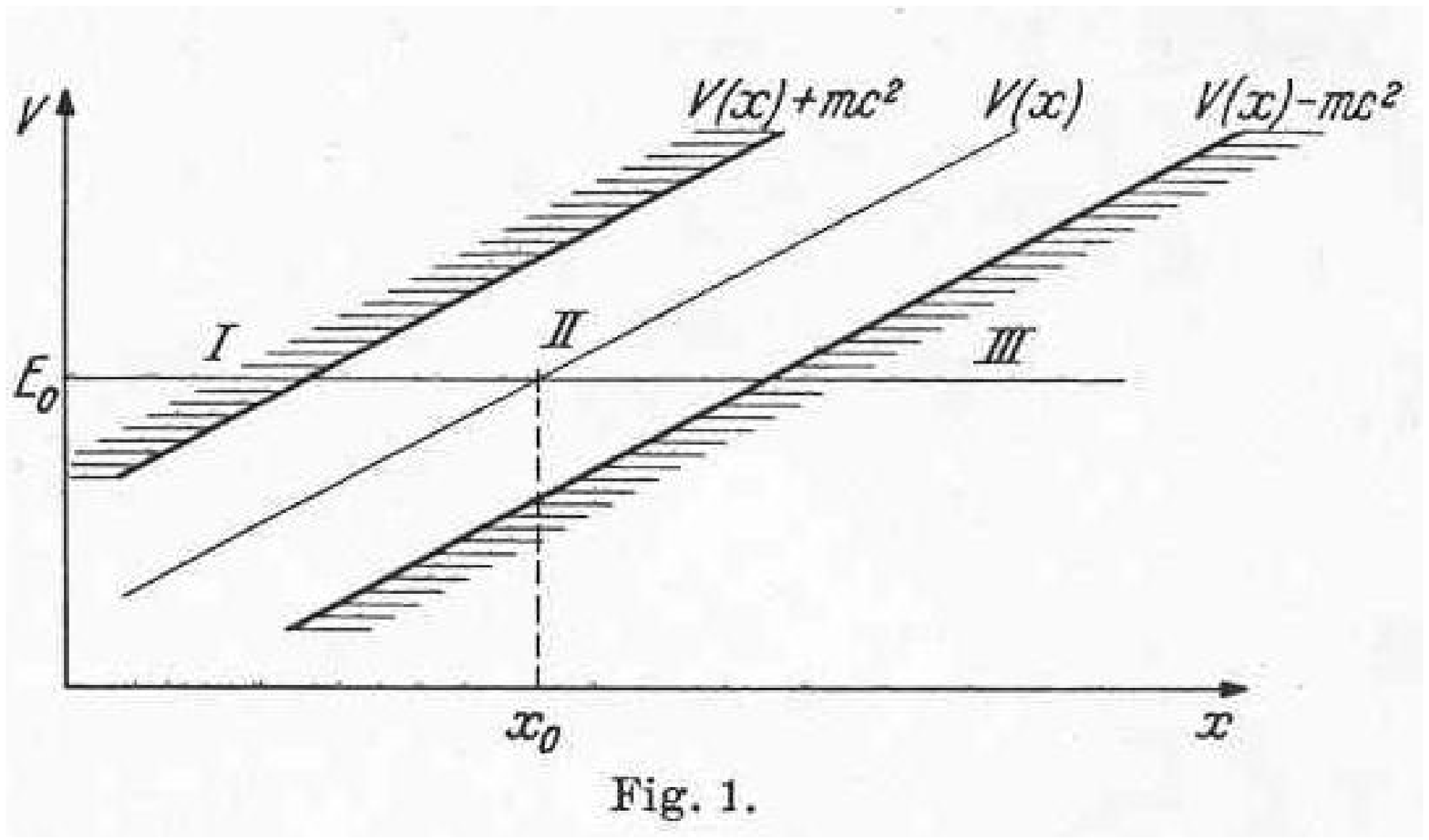}
\end{figure}

In Fig. 1, the potential energy $V(x)$ and the lines $V(x)+mc^2$
and $V(x)-mc^2$ are plotted against the
coordinate (the electric field is parallel $x$-axis).
The calculations of Sauter show that the eigenfunction associated to the eigenvalue
$E_0$, for example, is large only in the regions I and II. In the region II, they decrease exponentially.
Therefore, a wave function that begins being large in region I decreases slowly in region III where the
transmission coefficient through region II (which plays the role of a Gamow-wall)
calculated by Sauter has the order of magnitude $e^{-\frac{m^2c^3}{\hbar e |{\mathfrak E}|}\pi}$.
If we define $|{\mathfrak E}_k|=\frac{m^2c^3}{\hbar e}$ as the critical field strength,
we can also write $e^{-\frac{|{\mathfrak E}_k|}{|{\mathfrak E}|}\pi}$. As long as $|{\mathfrak E}|\ll|{\mathfrak E}_k|$,
pair creation is so rare that it can be practically ignored. Then it must be possible to find solutions
of the Dirac equation playing the role of eigenfunctions, which are large in region I but stay small
of the order of $e^{-\frac{|{\mathfrak E}_k|}{|{\mathfrak E}|}\pi}$ in region III.
Conversely, it must be possible to find solutions which are large in region III
and small in region I. After that we can characterize the state of the lowest energy by
all electron states being occupied whose eigenfunctions are large only in region III,
while the others are unoccupied. The energy
density at $x_0$ is calculated from
the differences of the electron energies
with $E_0$ [compare Eq. (31)].
By switching off the electric field adiabatically, the
so characterized state of the system goes over into the state of the field-free space, in which only the
negativ-energy electron states are occupied.

Our calculations follow those of Sauter. If an external magnetic field $\mathfrak B$
and an electric field $\mathfrak E$ are present, both pointing in $x$-direction, the Dirac equation reads:

 \includegraphics[scale=0.6]{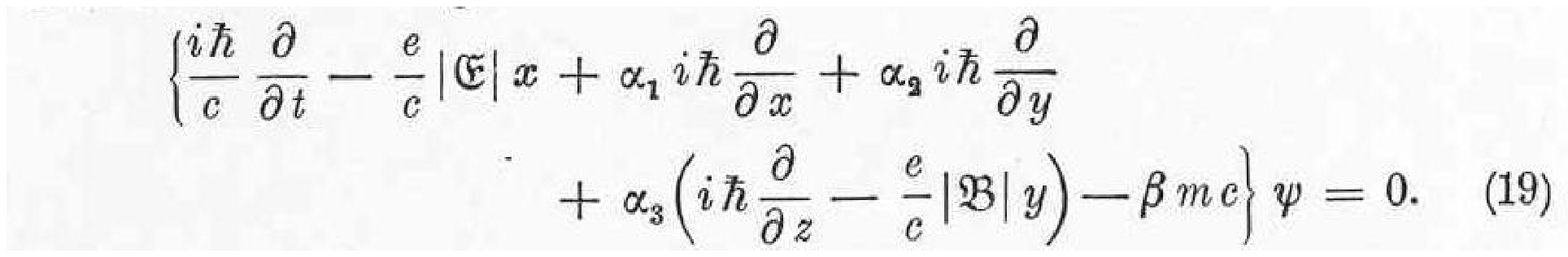}
The movement in $y$- and $z$-directions can be separated from that in $x$-direction
by the ansatz:

\includegraphics[scale=0.6]{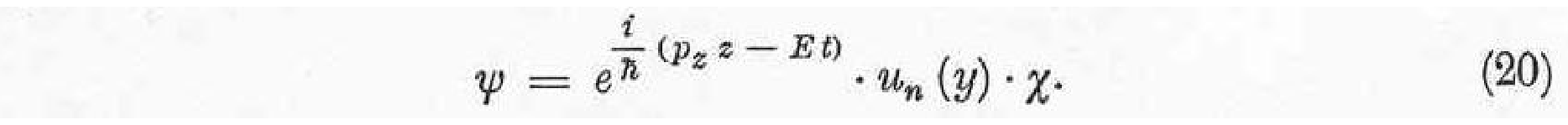}
We introduce a new operator $K$ by

 \includegraphics[scale=0.6]{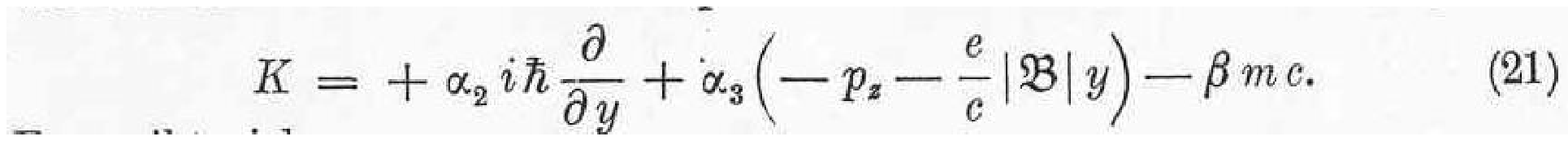}
It satisfies:

 \includegraphics[scale=0.6]{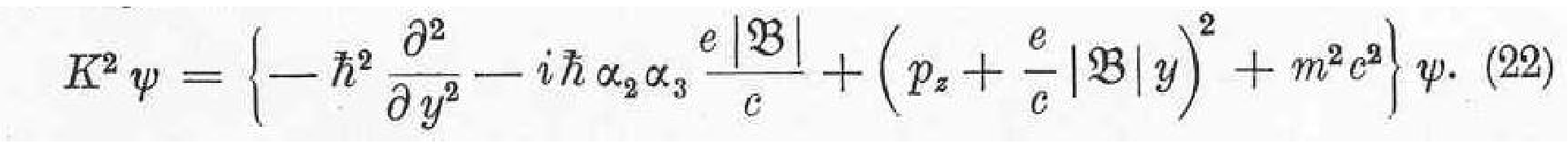}
This equation can be understood as a wave equation determing eigenfunctions $u_n(y)$.
If we set

 \includegraphics[scale=0.6]{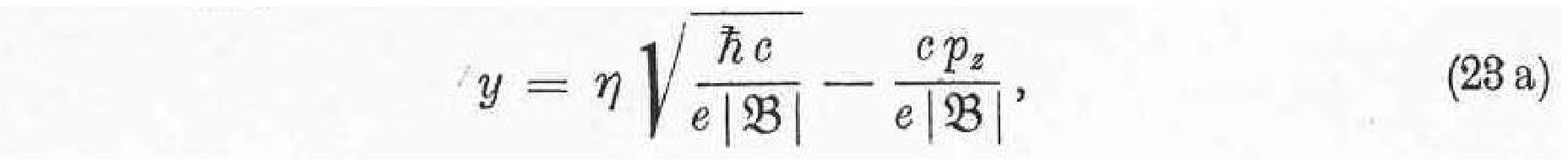}
we find (22) which becomes essentially the Schroedinger equation of the oszillator,

 \includegraphics[scale=0.6]{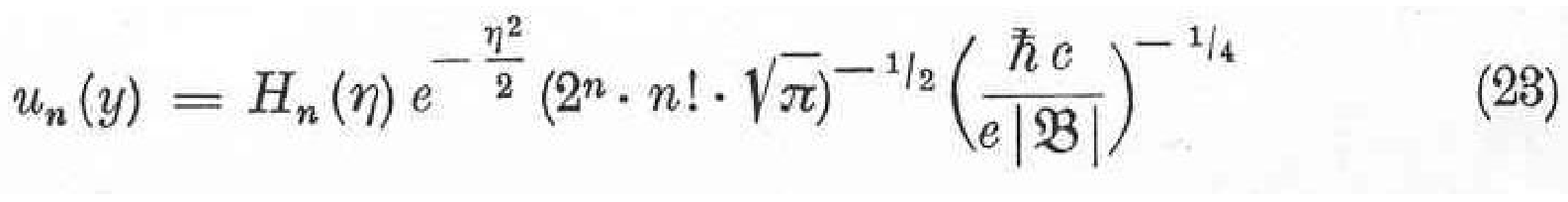}
($H_n(\eta)$ is the $n$-th Hermite polynom). The eigenvalues are:

\includegraphics[scale=0.6]{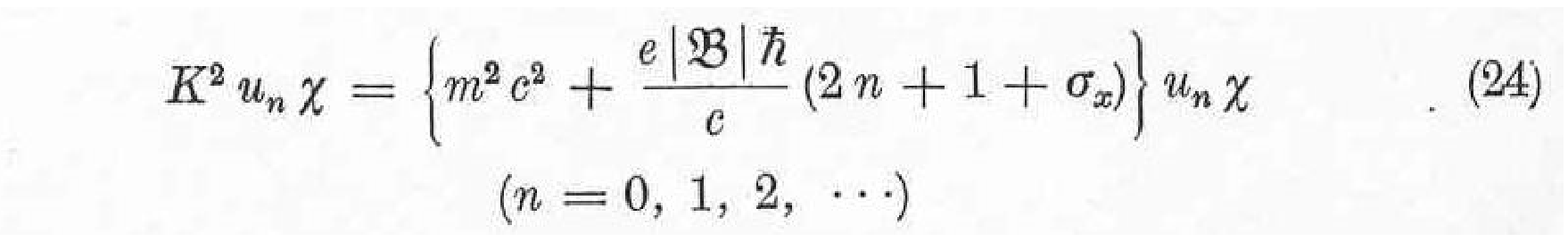}
The operator $K$ anticommutates with $\alpha_1$ in the wave function (19), which can also
be written as

\includegraphics[scale=0.6]{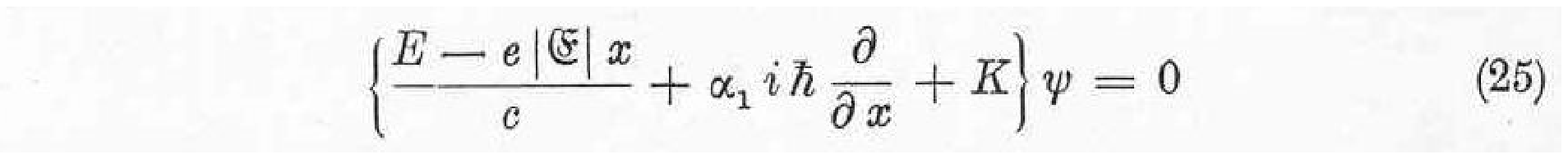}
We can perform a canonical transformation in $\chi$, so that $\sigma_x$ becomes diagonal,
and $K$ and $\alpha_1$ read:

\includegraphics[scale=0.6]{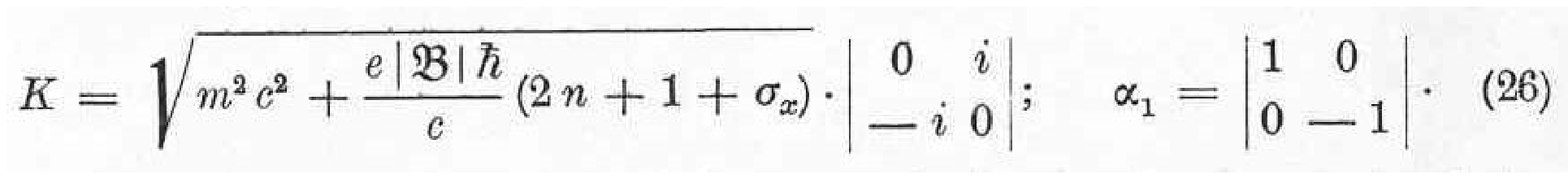}
Here the two matrices refer to another index independent of the spin orientation
(they refer to the ``$\varrho$''-coordinate). One can treat $\sigma_x$ as if it was
an ordinary number ($\sigma_x=\pm 1$) and obtains with the convenient shorthand notation

\includegraphics[scale=0.6]{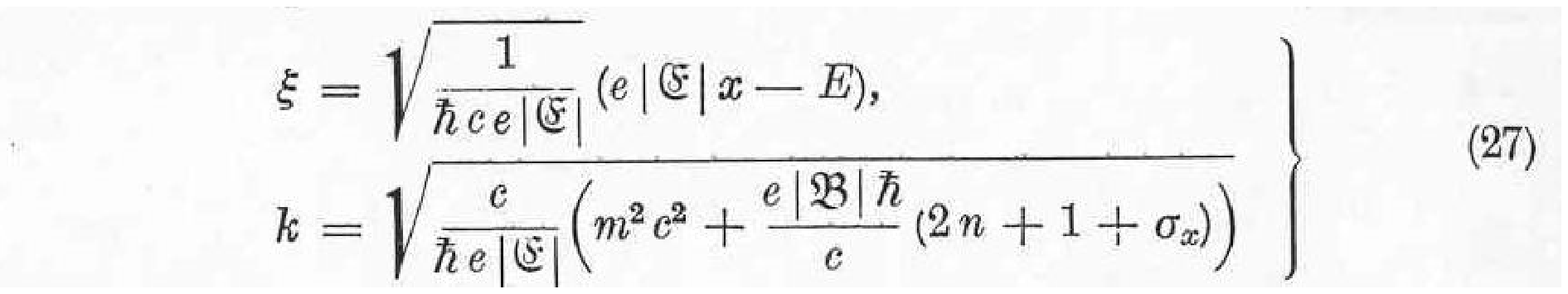}
the equations

\includegraphics[scale=0.6]{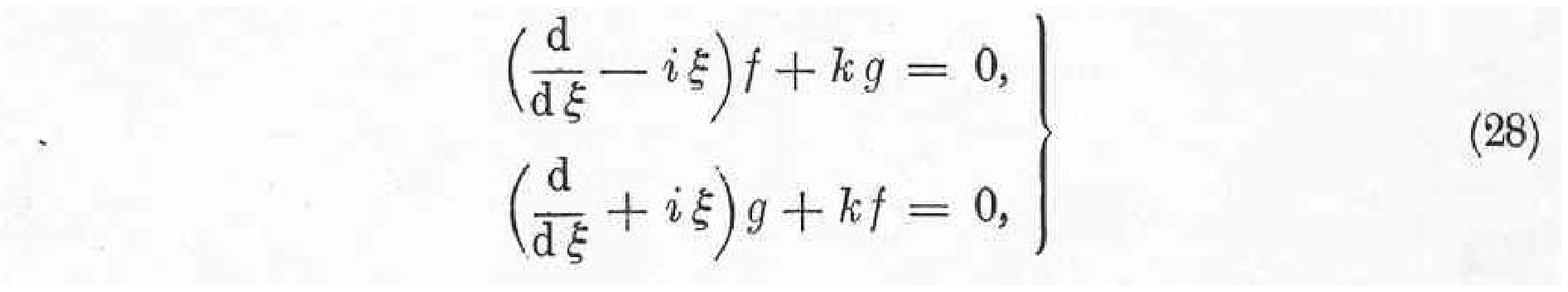}
where $f$ and $g$ are the two components of the function $\chi$ (associated with the $''\varrho''$-index).
The equations (28) are formally identical with the Equations (12) from Sauter''s paper (op. cit. \ref{Sauter}).
They differ from those, however, in the meaning of $k$ and that the system (28) actually has to
be written down twice; once for $\sigma_x=+1$
and once for $\sigma_x=-1$. Sauter obtains two sets of solutions:

\includegraphics[scale=0.6]{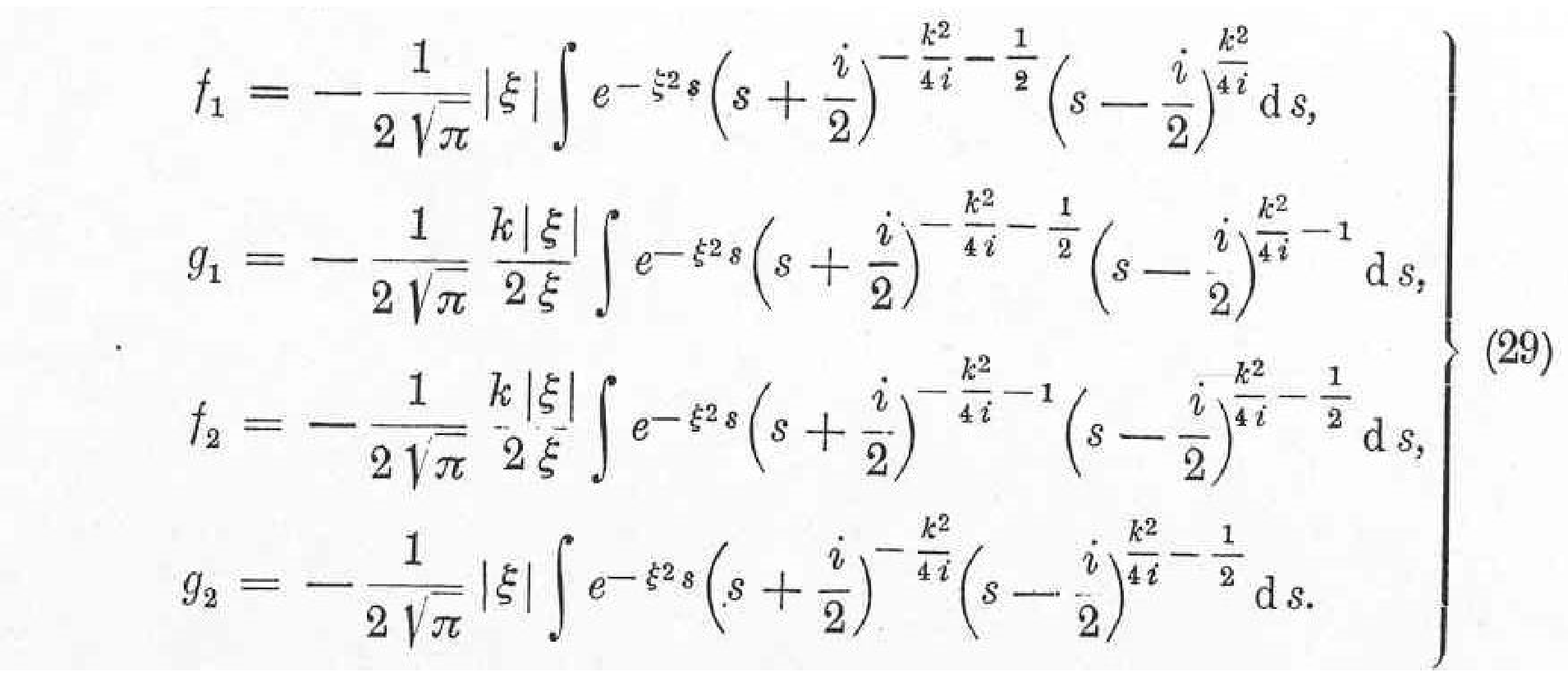}
The integrals are to be taken along a path coming from $+\infty$ circulating around both singular points
$+i/2$ and $-i/2$ in a positive manner, and returning to $+\infty$.

Since we ignore pair creation in the calculation we consider as eigenfunctions only those parts of the functions
$f$ and $g$ which vanish in one half of the space. Thus we set

        \includegraphics[scale=0.6]{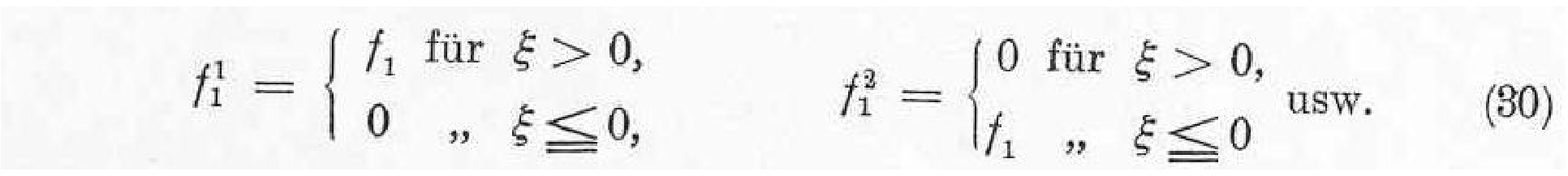}
The new functions $f^1_1$ etc. do not correspond exactly to the stationary states but
represent wave packets for which it is unlikly to diffuse into the initially empty
area. For the construction of the density matrix we consider the states $f^1_1$, $g^1_1$, ...
as being occupied, the states $f^2_1$, $g^2_1$, ... as unoccupied.
We have doubled the number of states by the process (30). Thus, if we take all $f^1_1$, $g^1_1$, $f^1_2$, $g^1_2$
as occupied and $f^2_1$, $g^2_1$, $f^2_2$, $g^2_2$ as unoccupied, we shall obtain the density
matrix twice.

To find the energy density of the vacuum according to the method described in the first section we
would have to calculate the density matrix associated with a
finite distance of the two positions $\mathfrak{r'}$
and $\mathfrak{r''}$. From this we have to subtract the singular matrix $S$.
Then obtain
the energy density by taking
the limit $\mathfrak{r'}=\mathfrak{r''}$. For the following calculation
it is more convenient to set
 from the beginning $\mathfrak{r'}=\mathfrak{r''}$, but to restrict the summation over
the stationary states to those of finite energy only. Aquivalently, we could make the summation convergent
by an auxiliary factor $e^{-\mathrm{const}[E^2-(mc^2)^2]}$.
If the constant number in the exponent goes to zero, some of the terms in the energy density
become singular, and they will be compensated by corresponding elements in the matrix $S$.
The remaining regular elements yield the desired result.

Bevor writing down the density matrix, we have to normalize the eigenfunctions.
We can imagine the space of the eigenfunctions in $x$- and $z$-direction being
confined to a large length $L$ (the above eigenfunctions $u_n(y)$ are already normalized).
Then we have from the $z$-direction a normalization factor $1/\sqrt{L}$; from the
$x$-direction, the asymptotic behavior of the Sauter eigenfunctions [compare Eq. (22)]
yields the factor $2\frac{1}{\sqrt{L}}e^{-\frac{k^2\pi}{4}}$. The summation over all
states is to be done over all momenta which have the form

$$p_z=\frac{h}{L}m+\mathrm{const}$$
and over all energies of the form $E=\frac{hc}{L/2}m+\mathrm{const}$.
These two sums can be converted into integrals. When neglecting the factor $\frac{1}{\sqrt{L}}$
the differential is $\frac{dp_x}{h}\frac{dE}{2hc}$. If we calculate the density at position
$x_0$, the energy of a state is the difference $E-e|{\mathfrak E}|x_0$. We therefore obtain for
the energy density corresponding to the matrix $R_S$ (compare Section 1) [for the meaning of $a$ compare (33)]

 \includegraphics[scale=0.6]{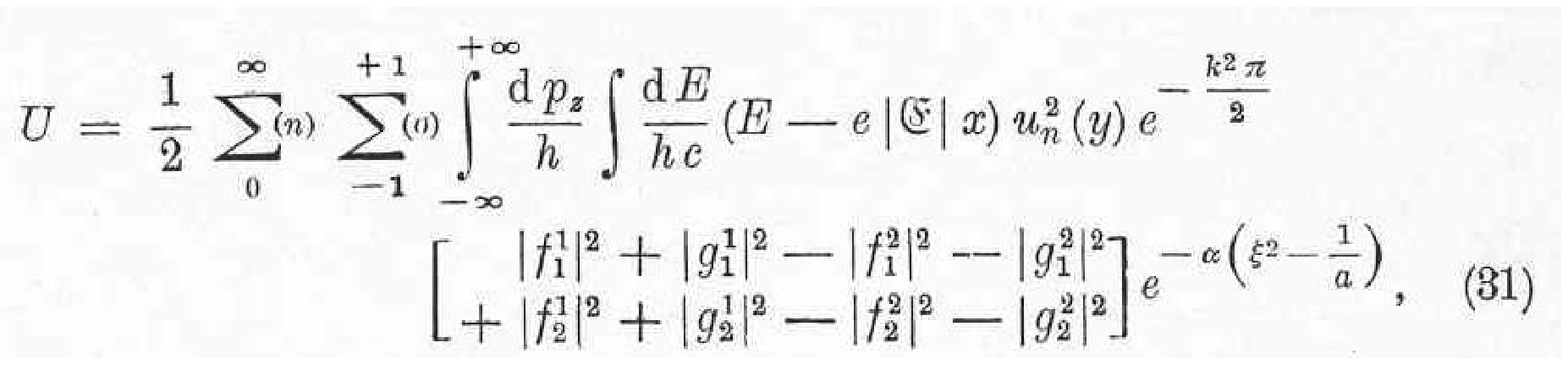}
which due to (23a) goes over into

        \includegraphics[scale=0.6]{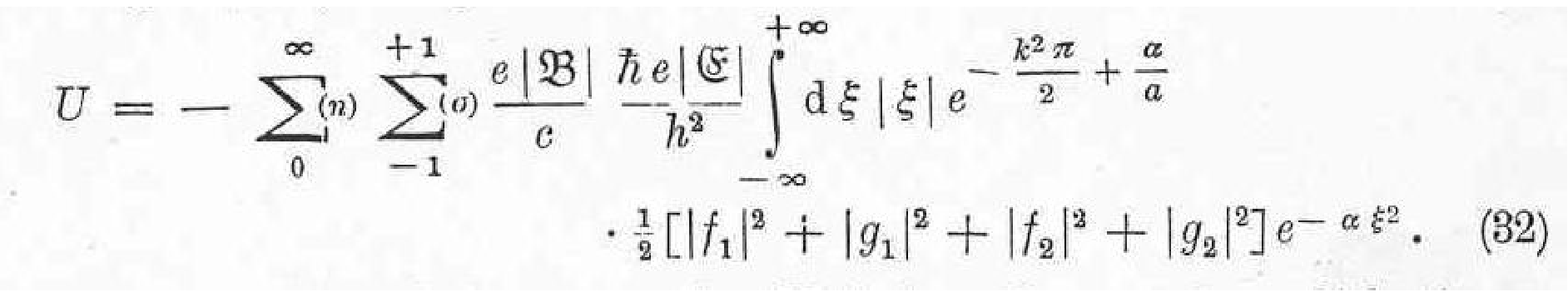}
From this expression we have to derive the limit $\alpha\overrightarrow{}0$.
We introduce the following convenient abbreviations:

        \includegraphics[scale=0.6]{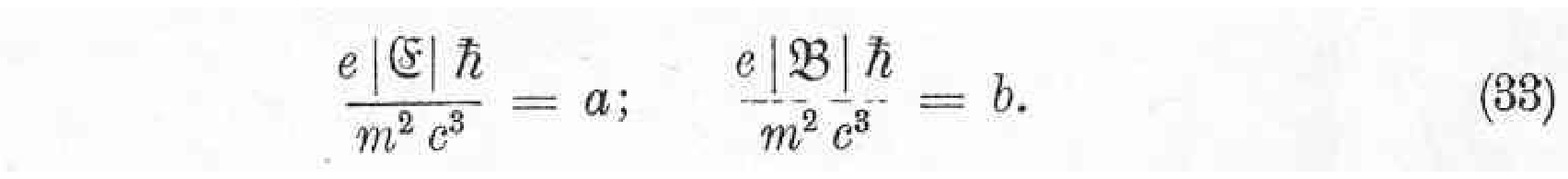}
These are dimensionsless and represent the ratio of the field strengths with the critical field strength
$ |\mathfrak {E} _k | $, i.e., with the ``$1/137 $th fraction of the field strength at the radius of the electron''.
With the equation (29) one finally finds

\includegraphics[scale=0.6]{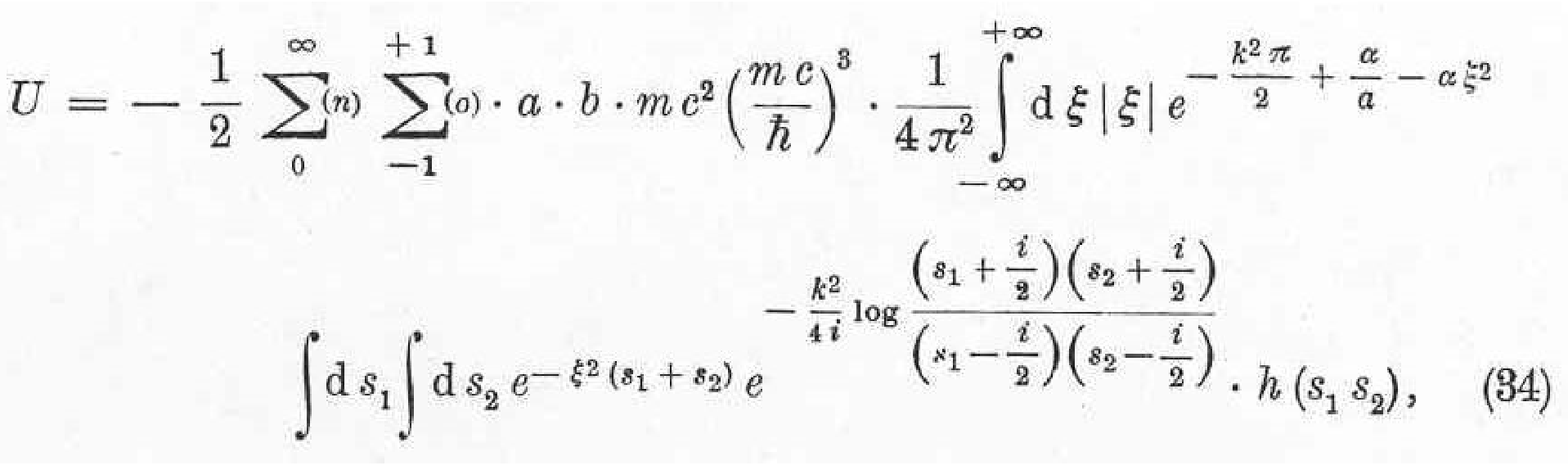}
where $h(s_1 s_2)$ is

\includegraphics[scale=0.6]{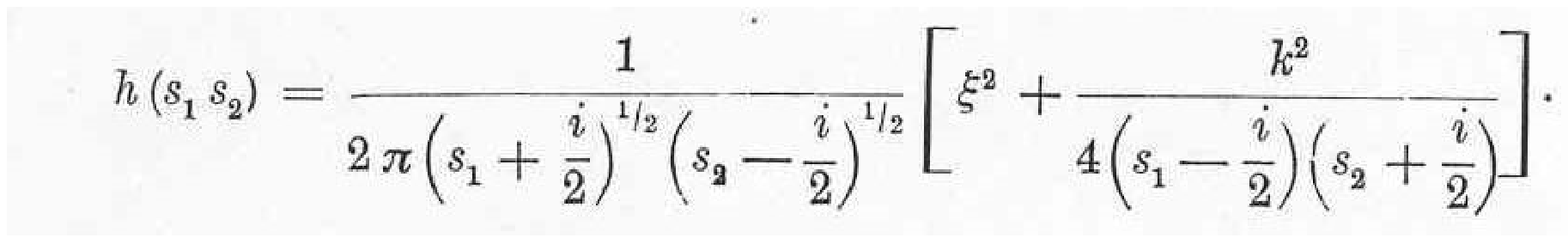}
The integration over $\xi $ yields:

\includegraphics[scale=0.6]{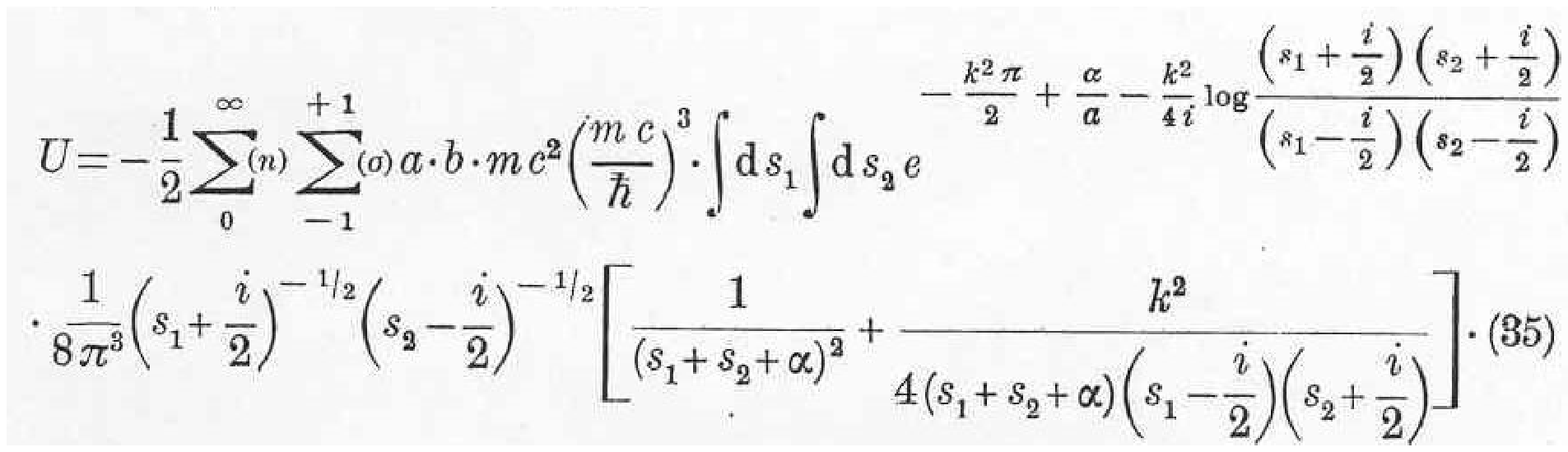}
The first of the two $s$-integrals, i.e., the one over $s_1$, can be performed by deforming the integration
path in such a manner that only a loop around the pole $s_1 =s_2 -\alpha $ is left. Replacing $s_2 $ by $ s=s_2 - \alpha $
in the result, one obtains:\\

\includegraphics[scale=0.6]{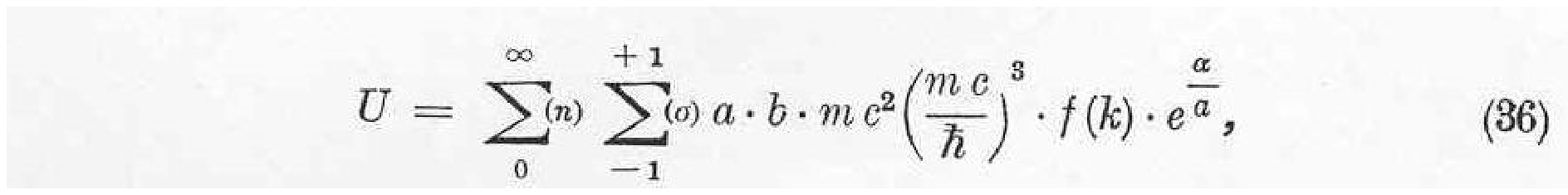}
where

\includegraphics[scale=0.6]{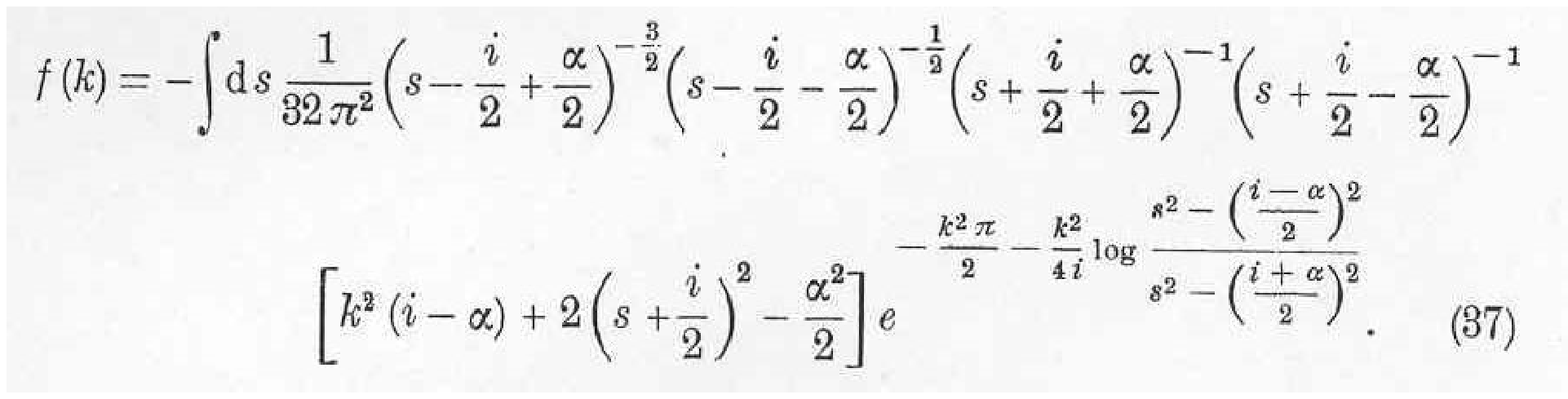}
The integration is to be carried out along a path shown in Fig. 2. The contour comes from $+\infty $ passes
between the four poles, and returns back to $+\infty $. Alternatively, one may integrate along the imaginary axis
from $+\infty $ to $-\infty $.

\begin{figure}[h]
\centering
\includegraphics[scale=0.3]{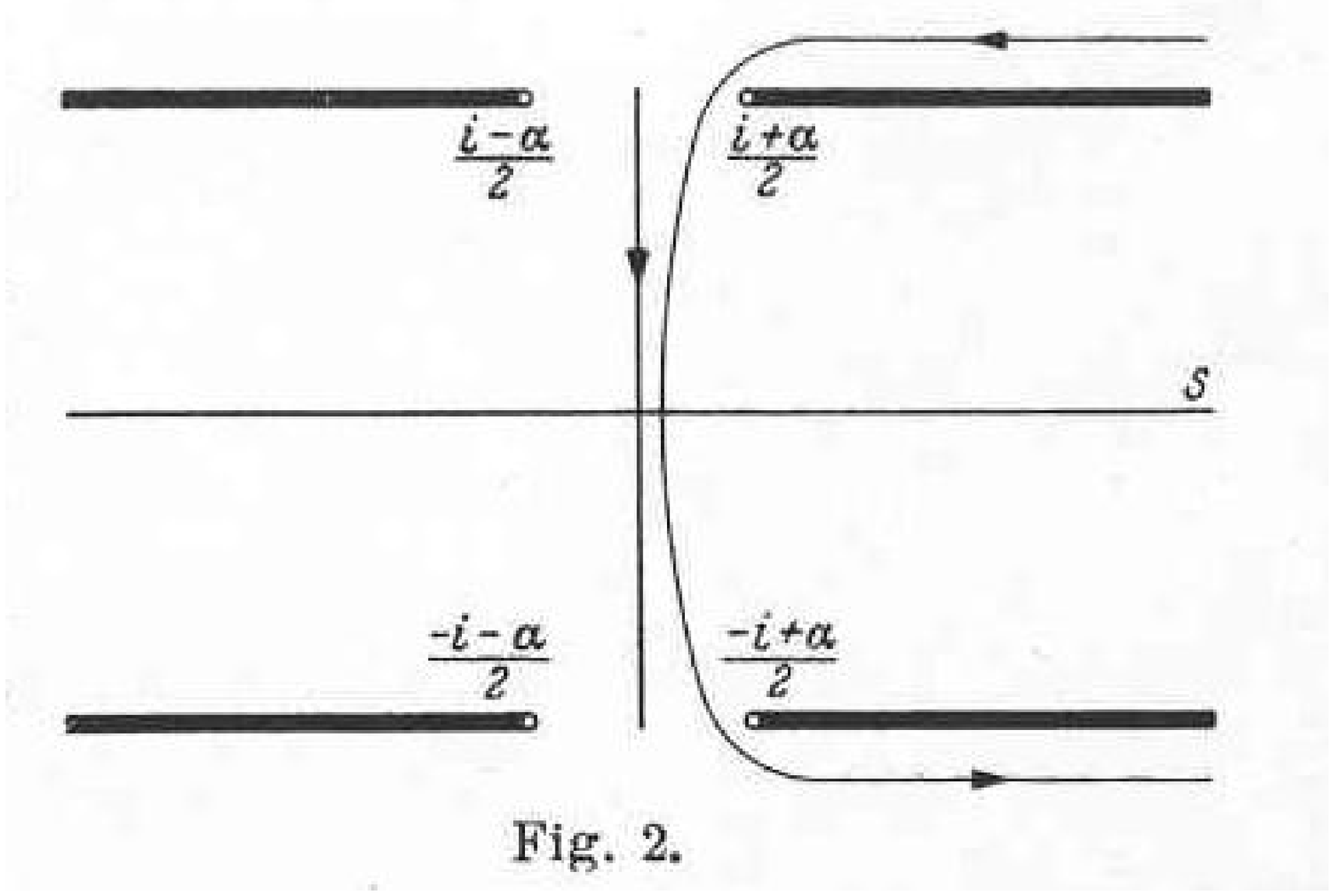}
\end{figure}
\pagebreak

The main contribution to the integral comes from the regions between the poles. One may expand the logarithm
in the exponent in power series of $ \alpha $, and obtains~\\

\includegraphics[scale=0.6]{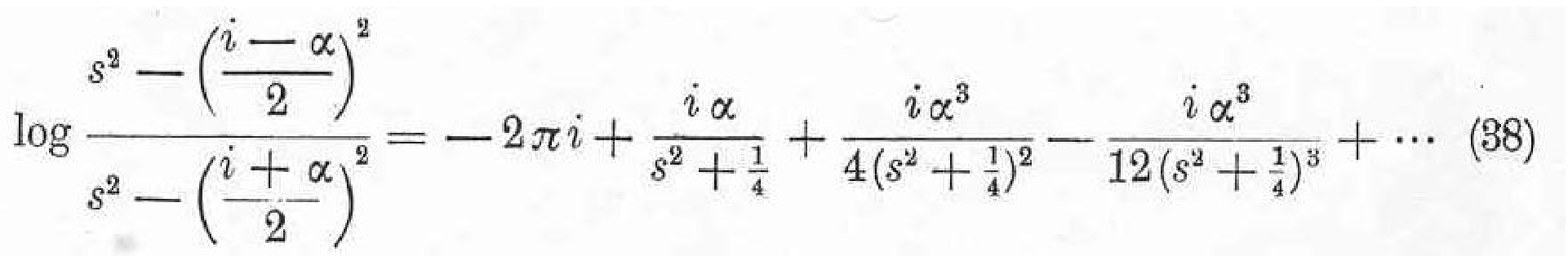}                  ~\\
For the following calculations, it is assumed that the electric field is small against the critical field
$|\mathfrak {E} _k | $, i.e. $a\ll 1 $ and thus \\

\includegraphics[scale=0.6]{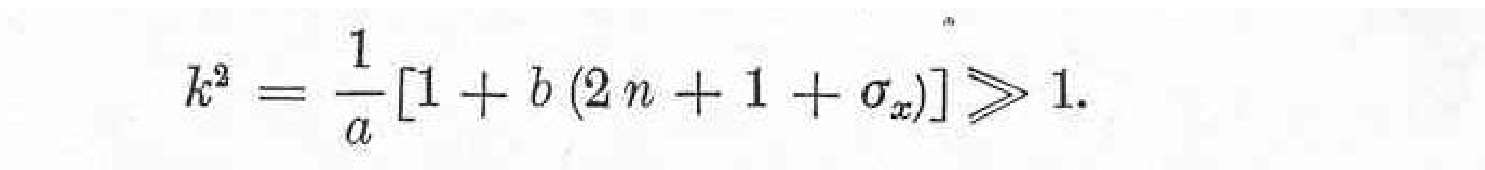}      ~\\
Then the expression $-\frac{k^2 \alpha}{4s^2 +1}$ in the exponent must be retained; the higher terms in the exponent
 may be considered as small, while we shall eventually take $\alpha\overrightarrow{}0$. They can therefore be included
by an expansion. Thus, one obtains an expression for $f(k)$ of the form \\[5mm]

\includegraphics[scale=0.6]{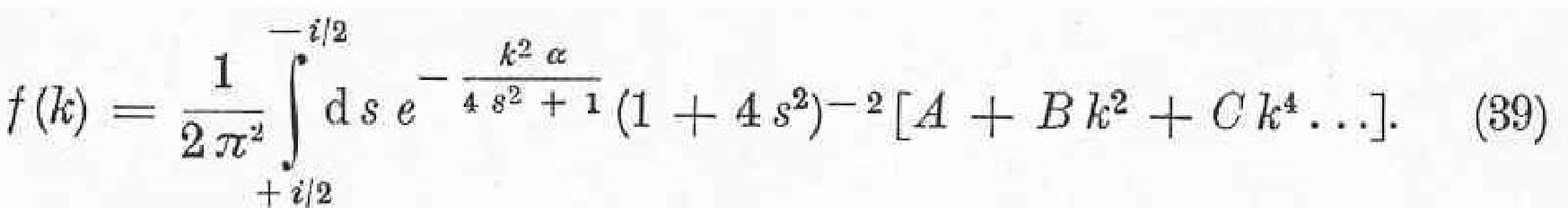} \\[5mm]

Before performing the integral, it is convenient to calculate the summation over $n $ and $\sigma $; it is carried
out according to the scheme                   \\

\includegraphics[scale=0.6]{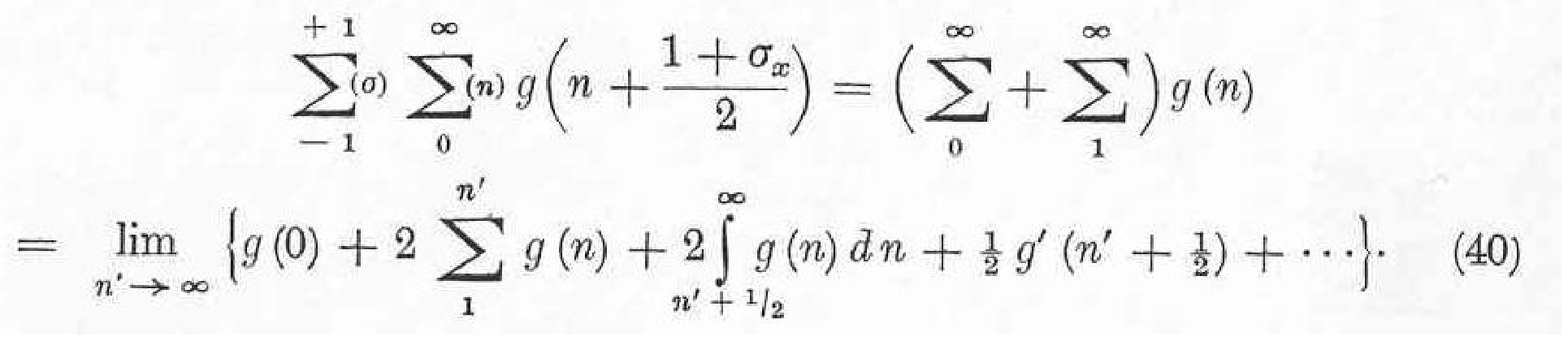}     \\~\\
It turns out that the higher terms of Euler summation formula contribute nothing to the end result. Finally,
setting $\frac{\alpha}{a}=\varepsilon $, one gets for $\varepsilon \to 0$ (where $\gamma$ denotes the Euler constant $ \gamma = 1.781$).

\pagebreak
\includegraphics[scale=0.6]{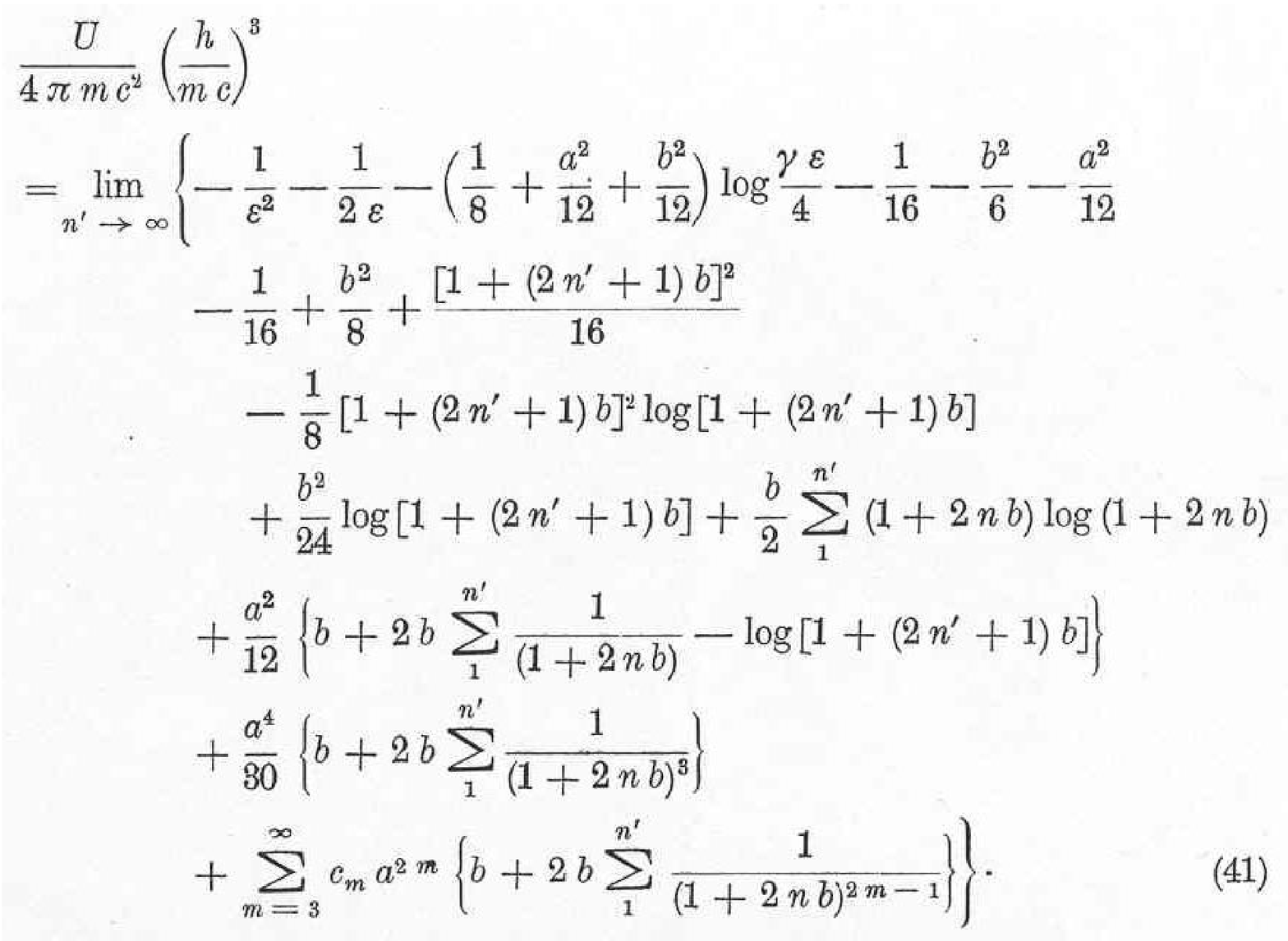}
(The coefficients $c_m $ are to be determined further below).

The parts corresponding to the singular matrix $ S $ still have to be subtracted from the result. One obtains
easily the field-independent part of the singular energy density, which is to subtract by repeating the above
calculation with plane waves:

\includegraphics[scale=0.6]{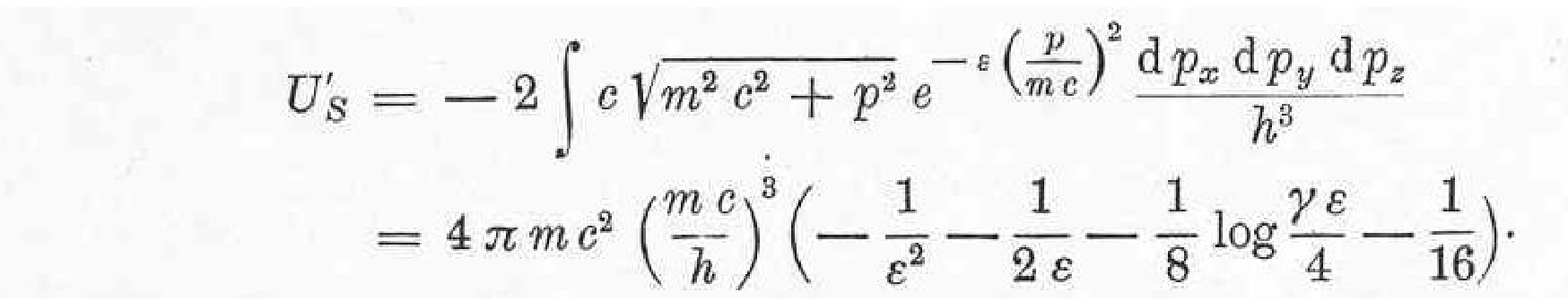}
It is more difficult to calculate the field-dependent part of $S$. Since equation (13) contains only squares of
field strengths in $\bar{a}$ and $\bar{b}$, this would also for $U_S $. Further, the constant C in equation
(13) is chosen so that for constant fields, polarisation proportional to the field does not
occur. From this it follows that the terms which are quadratic in the field are to be subtracted, and only the
higher terms remain. This implies, anticipating an expansion in powers of $b$ for $b\ll 1 $, that the first line of
the right-hand equation (41) is to subtract on the whole. We proved this result by assuming that the points
$ \mathfrak {r} '$ and $\mathfrak {r} ''$ of the density matrix are different in the $x$ direction
$ [\mathfrak {r} ' -\mathfrak{r} ''=(x,0,0)]$, and by setting at the end $\alpha =\epsilon =0 $. After subtracting the terms
which come from the matrix $ S $, only the above discussed part of $U $ remains. For the electric
field, this calculation turned out to be very difficult. Starting with the total energy density, which is
composed of the usual Maxwell energy density $ \frac{1}{8\pi}(\mathfrak{E} ^2 +\mathfrak{B} ^2) $ and
Dirac energy density $ U-U_S $,
we go on with the Lagrangian using the relation (3)

\includegraphics[scale=0.6]{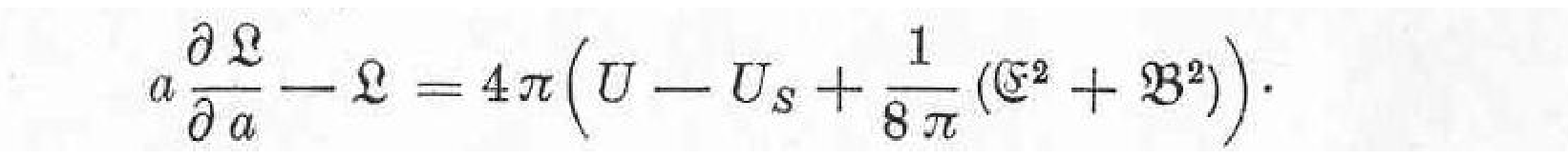}
One obtains

\includegraphics[scale=0.6]{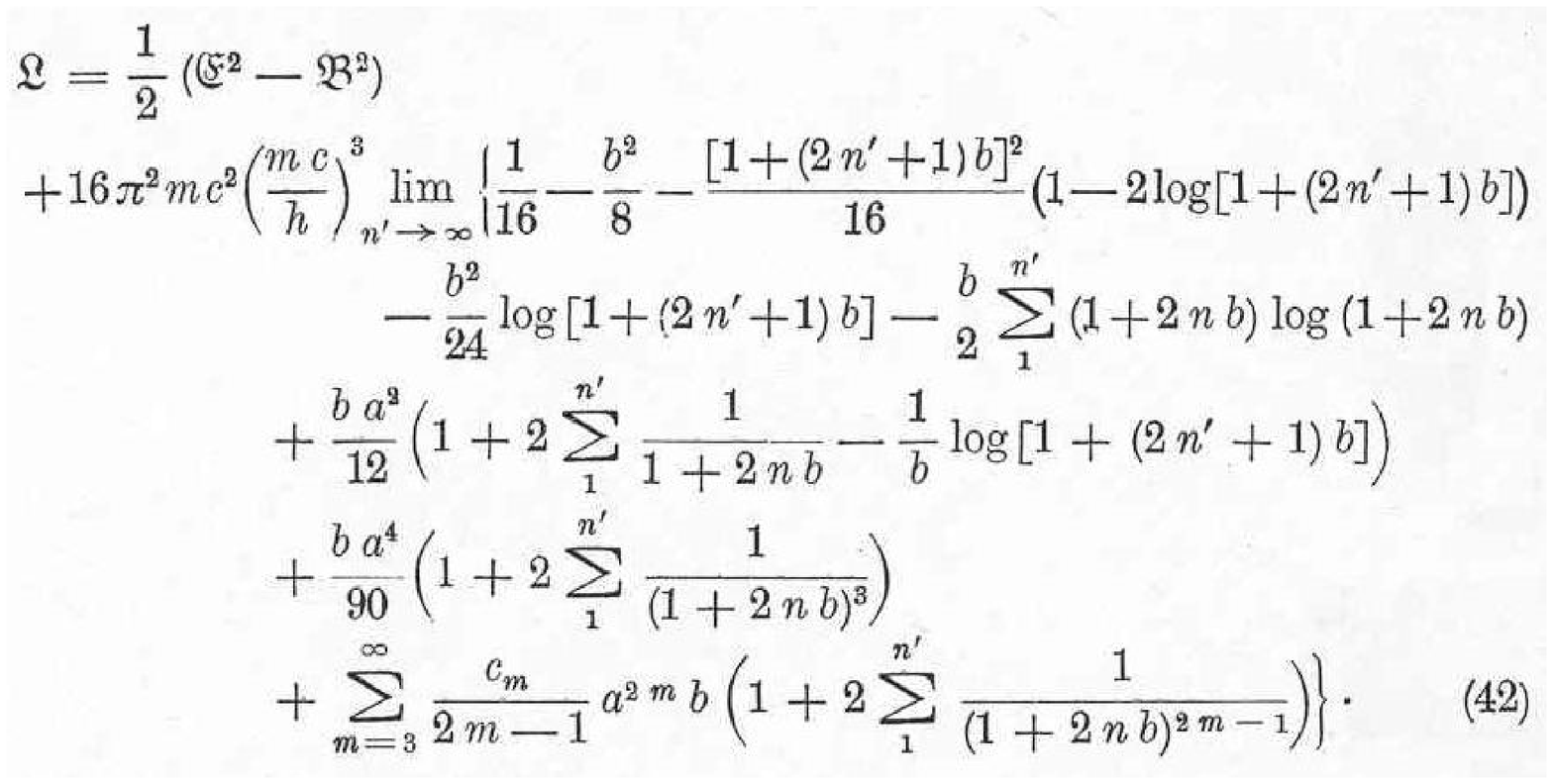}
For small magnetic fields, the power series expansion of $ b $ may be found applying to equation (42) once more
the Euler summation formula. Furthermore, since $ \mathfrak{L}$ depends only on both invariants $ a^2 -b^2 $ and $ a^2 b^2 $, from which it follows that $ \mathfrak{L} (a, b)$: $ \mathfrak{L} (a, 0)= \mathfrak{L} (0, ia) $, the missing
coefficients $ c_m $, whose direct calculation would be very involved, may be determined indirectly from this relation. By calculating $ c_2 $ and $c_3 $ we have checked that the direct calculation of $ c_m $ yields the same result; though we didn't find the general proof for it. In this manner, one obtains initially for small fields ($ a \ll 1$, $b \ll 1$):

\includegraphics[scale=0.6]{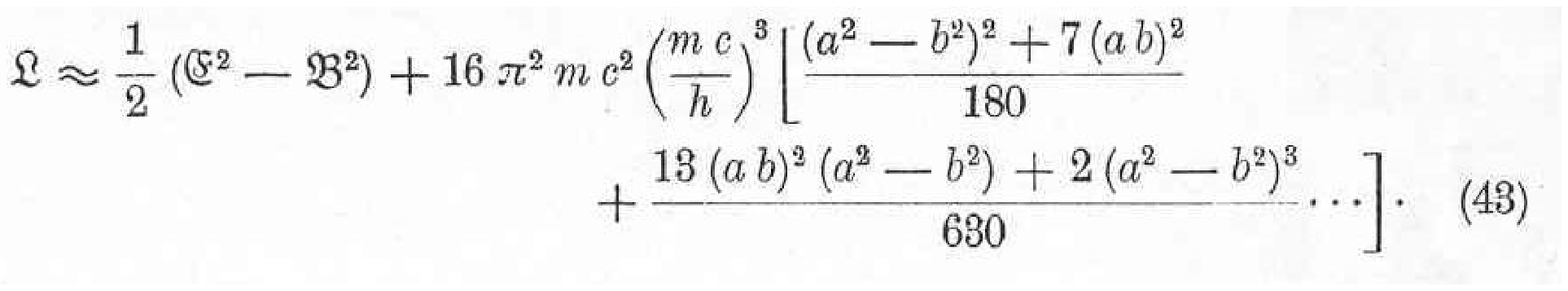}
The opposite limit ($ a \ll 1$, $b \gg 1 $) yields

 \includegraphics[scale=0.6]{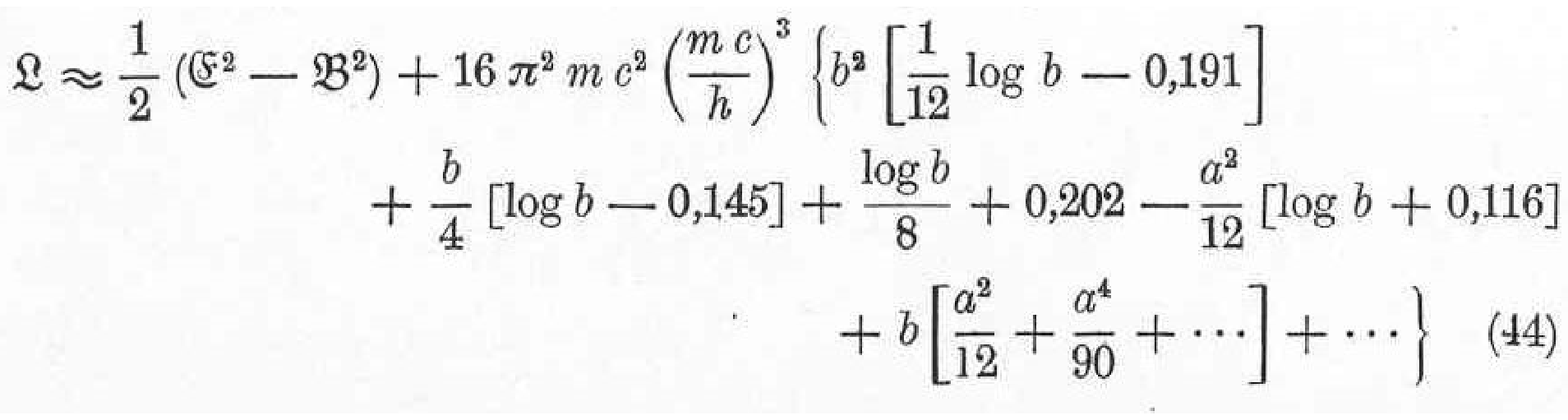}
We tried to derive an integral representation of $ \mathfrak{L}$ in order to better see the general behavior of
$ \mathfrak{L}$ for an arbitrary field. This is possible when using the usual integral representation of the
zeta function. One obtains

  \includegraphics[scale=0.6]{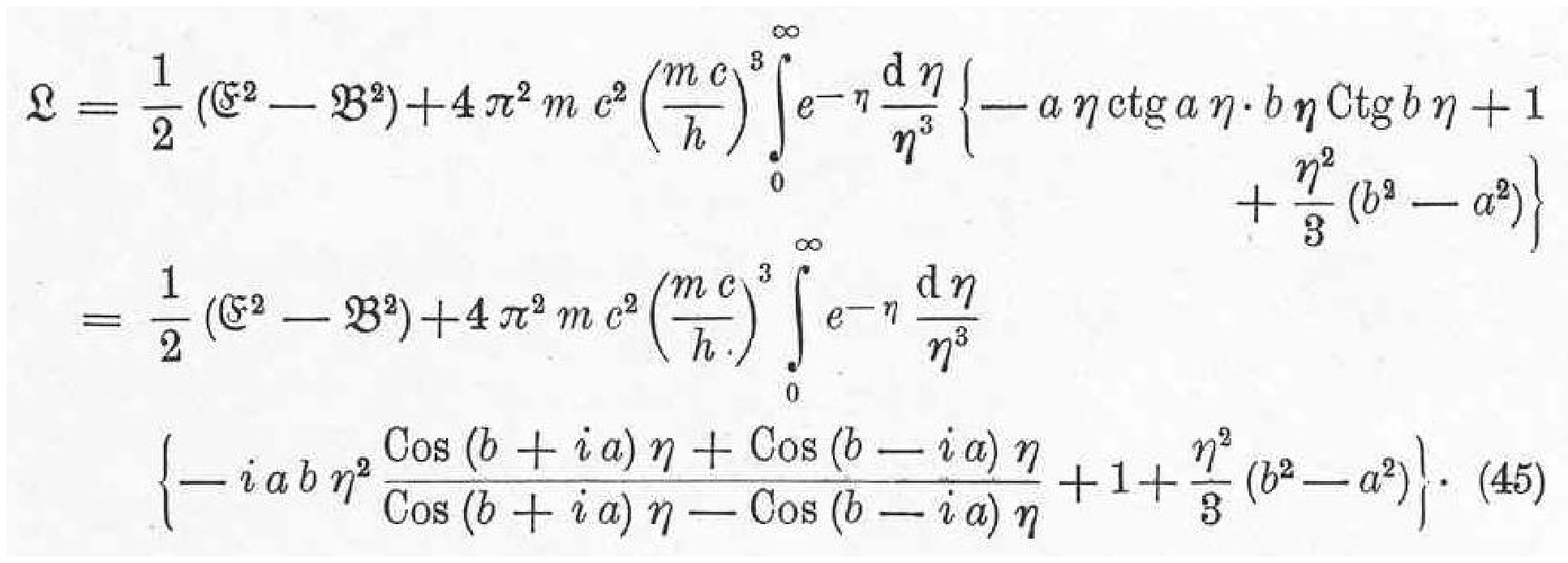}
From the last form it is easy to see that $ \mathfrak {L} $ only depends on both invariants $ \mathfrak {E} ^2
- \mathfrak {B} ^2 $ and $ (\mathfrak {EB})^2 $. The cos terms allow an expansion in squares of the argument
$ (b +ia)^2 = b^2 - a^2 + 2i(ab) $ and $ (b +ia)^2 = b^2 - a^2 - 2i(ab) $. Since the overall result is real,
it may be represented as power series in $ b^2 - a^2 $ and $ (ab)^2 $, which may be replaced by
$\frac{\mathfrak{B} ^2 -\mathfrak {E} ^2}{|\mathfrak {E} _k | ^2} $ and
$\frac{\mathfrak{EB} ^2 }{|\mathfrak {E} _k | ^4}  $ respectively.
$\biggl( |\mathfrak {E} _k | ^2 = \frac {m^2 c^3}{e \hbar} \biggr) $.
Thus the Lagrangian for arbitrarily oriented fields reads:

\includegraphics[scale=0.6]{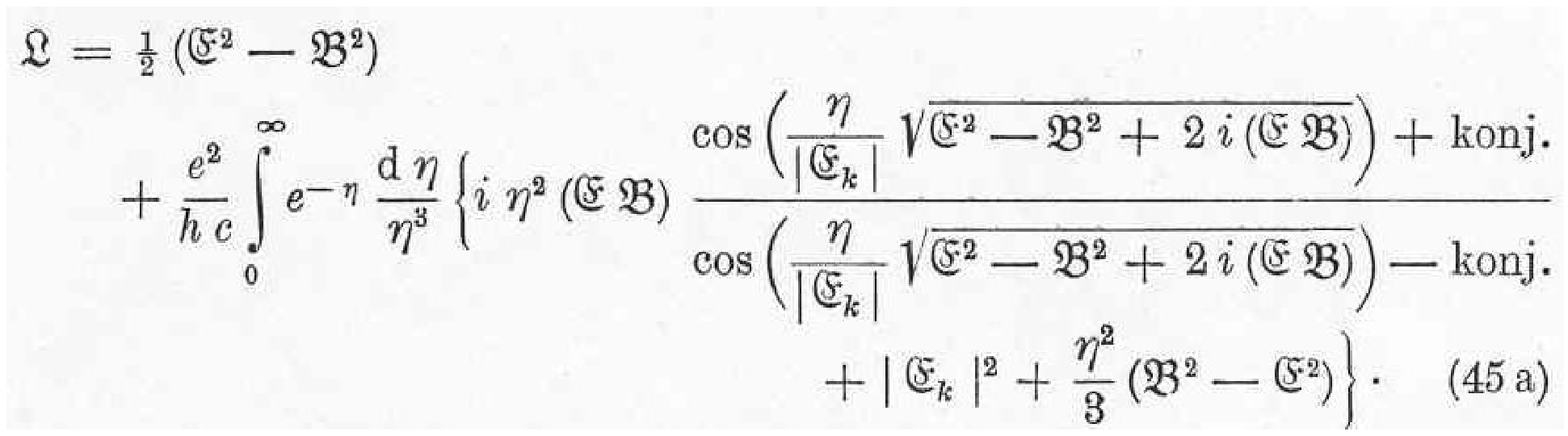}
The first expansion term of equation (43) agrees with the results of Euler and Kockel (op. cit. \ref{EulerKockel}).

The question of convergence of this power series expansion deserves closer investigation. The integral (45)
converges for every value of $ b$ when $ a=0 $. But when $ a \neq 0 $, the integral looses its meaning for
$ \eta = \frac{\pi }{a},\,  \frac{2 \pi }{a}, \cdots $, where $\cot a\eta $ becomes infinite. According to this, the power
series expansion in $ a $, from which we started, can only be semiconvergent. We can give a definite meaning
to the integral (45) by selecting an integration path that avoids the singular values
$ \frac{\pi }{a},\,  \frac{2 \pi }{a} $. But then the integral (45) acquires additional imaginary terms
that can't be interpreted physically. Their meaning becomes clear if we estimate their size.
The integral (45) around the pole $ \eta = \frac {\pi }{a} $ has the value
$ -\frac {2i}{\pi }\cdot 4 a^2 mc^2 (\frac {mc}{h})^3 \cdot e^{-\frac{\pi }{a}}$ (for $ b=0 $). This is the order of the terms
which are associated with the pair creation in the electric field. Thus, the integral is to be interpreted
similar to the integration of the resonance denominator in perturbation theory. One may assume that a damping
term corresponding to the frequency of the resonance process is responsible for the convergence. Further, one
may assume that the result, which is obtained by avoiding the singular values, is correct except for the terms
whose order corresponds to the frequency of the resonance process.

The deviations from the Maxwell theory remain very small according to (43) and (44) as long as $ \mathfrak{E} $
and $ \mathfrak{B} $ are small against the electric field present at a distance of
$ \sqrt{137} \cdot \frac{e^2}{mc^2} $ from the center of the electron. But also in the case that the magnetic field
exceeds this value, the additional terms to the Maxwell equations remain small (proportional to the original terms).
They have the relative order $ \frac {1}{3\pi} \frac{e^2}{\hbar c}$ as long as $ \log {b} $ is near unity. Thus,
the deviations from the usual Coulomb force between the protons (for example) come from the terms (43) and (44),
and remain always very small. However, in this estimate, one must take into account that for a Coulomb field, the
neglected terms containing the derivatives of the field strengths are more important than those contained
in equations (43) and (44).

\section{Importance of the Result in the Quantum Theory of Wave Fields}
The results derived in the last section may not be transfered immediately to the quantum theory of wave fields.
In fact, one may show that the state of matter in a homogeneous field in the quantum theory of wave fields is not
described by the above equations.
Even if we start with the state of matter discussed in the last section as an ``undisturbed state'', there are
matrix elements of the perturbation theory which belong to the simultaneous formation of a light quantum and a pair.
Even, if the energy does not suffice for a real formation of these particles, such matrix elements give rise to a
perturbative energy of second order. This energy comes about through the virtual
possibility of formation and disappearance of the light quantum and a pair, and the calculation provides a
divergent result for them. The appearance of these perturbation terms may be made plausible by observing that
the circular trajectories in a magnetic field are not really stationary states, but that the electrons in these
states can radiate. In the classical theory of wave fields, this radiation can be ignored - this being crucial
for the physical content of the calculation of the last section. Indeed, in the final solution the charge density
and current density of matter vanish, so that radiation does not occur. In the quantum theory of wave fields
on the other hand, this radiation does appear in form of the divergent perturbative energy of second order.

One may also understand that a perturbative energy of exactly the same type occurs also in the field free vacuum
(``self-energy of the vacuum''). Such self-energies are found always when calculating the contribution
of second order or higher to the energy, which arises from virtual transition to another state and return to
the initial state. These self-energies have so far always been ignored. For instance, the cross
section of the
Compton scattering may be calculated by performing a perturbation calculation to second order. Considering
the terms of the forth order, one would get a contribution of the described type, which does not yield a
convergent result. The calculation of the light-light scattering is carried out to the forth order
(the lowest order that yields a contribution to the relevant process). The contributions of the sixth order
would already diverge. The previous success of these calculations, for instance the Klein-Nishina formula, seems
to show that omitting the divergent contributions of the higher order leads to the correct results. If this is
the case, then it follows from the above discussion that the results of Section 2 can be transferred to the
quantum theory of wave fields. This is also physically plausible, since the occurrence of the above radiation
terms would remain incomprehensible according to the correspondence principle. Each single term in the expansion
of the energy density in powers of $ \mathfrak{E}$ and $ \mathfrak{B}$ can concretely be associated with a
scattering process, whose cross section is determined by it. The terms of forth order, for instance, stand for
the usual light-light scattering, the terms of the sixth order determine the cross
section of the process
whereby three light quanta are scattered on each other etc. Irrespective of the question whether it is physically
acceptable to neglect higher order terms, each expansion term in the result of the last
section agree with a direct
calculation of the corresponding scattering process in the quantum theory of wave fields if the perturbation
calculation is only performed to the lowest order that yields a contribution to the corresponding process.
In both calculations, the contributions of the terms are neglected which correspond to the formation and
disappearance of the light quantum and a pair. [The agreement of the terms of forth order with the terms obtained
by the direct calculation of light-light scattering (compare \ref{EulerKockel}) is therefore a test for the
correctness of the calculation.]
For this reason, it is not ruled out that even the results for $|\mathfrak{B}| \geqq  |\mathfrak{E} _k| $
can be applied to experience. This is, however, certainly not true for $|\mathfrak{E}| \gtrsim  |\mathfrak{E} _k| $;
The pair creation destroys the basis for the above calculations when the electric fields are large.

\section{The Physical Consequences of the Result}
The results derived in the second section are very similar to Born's%
\footnote{M. Born, Proc. Roy. Soc. London (A) {\bfseries 143 }, 410, 1933;
M. Born and L. Infeld, Proc. Roy. Soc. London (A) {\bfseries 144 }, 425, 1934; {\bfseries 147}, 522, 1934;
{\bfseries 150}, 141, 1935.}
derivation of Maxwell equations.
Also Born obtains a more complicated function of the two invariants, $\mathfrak{E} ^2 -\mathfrak{B}^2 $
and $(\mathfrak{EB}) ^2$ in addition to the classical Lagrangian $\mathfrak{E} ^2 -\mathfrak{B}^2 $.
Incidentally, because of the actual value of $\frac{e ^2}{\hbar c} $, this function agrees with (43) in the order of
magnitude of the lowest expansion terms (compare \ref{EulerKockel}). Though it is also important to emphasize
the differences of both results. Born has chosen the modified Maxwell equation as the starting point of the theory,
whereas this change in the Dirac theory is a really indirect consequence of the virtual possibility of the pair
creation. In addition, Dirac's theory predicts also terms, which contain higher powers of the field strengths (compare
\ref{Ueh}, \ref{Serber} ).
Thus, especially the question of the self-energy of the electrons cannot be decided only with the help
of these changes. The result of Born's theory imply that the changes of the Maxwell equations calculated here
suffice to remove difficulties with the infinite self energy is an important indication of the further
development of the theory.

In this context, we must also ask the question whether the results about light-light scattering etc. derived
from the Dirac theory may be considered to be final or whether it is expected that a future theory will lead to
another results. Without doubt, the theory of the positron and the present quantum electrodynamics must be
considered as temporary. Especially, the rules of finding the $S$ matrix seem to be arbitrary in this theory
(the inhomogeneity of Dirac's equation). The only reason for the matrix $S$ introduced in [\ref{Heisenberg}]
is the relative mathematical simplicity (together with some postulates on the formulation of the conservation laws).
Thus, a deviation of the future theory from the present theory may well be possible. Since such variations may have a
big influence on the Maxwell equations, the present theory is only reliable in its order of magnitude
and the qualitative form of these changes. Up to now, it is hardly possible to make definite statements as to
 the final form of the Maxwell equations in future theory, since it is essential to consider all the processes of
particles with very high energy (for instance, the occurrence of ``showers'').

%

\end{document}